\documentclass[publish]{IEEEtran}
\usepackage{amssymb}
\usepackage{amsmath}
\usepackage{mdwtab}
\usepackage{epstopdf}
\usepackage{multirow}

\usepackage{url}
\usepackage[breaklinks]{hyperref}
\usepackage[hyphenbreaks]{breakurl}
\usepackage{cite}
\usepackage{setspace}
\usepackage{enumerate}
\usepackage[dvips]{graphicx}
\usepackage{float}
\usepackage{array}
\usepackage{placeins}
\usepackage[ppl]{mathcomp}
\usepackage{booktabs}
\DeclareGraphicsExtensions{.eps}
\graphicspath{{c:/users/saeid/dropbox/Assignments/Journal/accepted/5-Augmentation/images/},{e:/dropbox/Assignments/Journal/accepted/5-Augmentation/images/}}
\usepackage{tabularx}
\usepackage{multirow}
\usepackage{sidecap}
\usepackage{wrapfig}
\usepackage{bigstrut}

\begin{document}
\title {Cloud-Based Augmentation for Mobile Devices: Motivation, Taxonomies, and Open Challenges}
\author{Saeid Abolfazli,~\IEEEmembership{Member,~IEEE,} Zohreh Sanaei,~\IEEEmembership{Member,~IEEE,} Ejaz Ahmed,~\IEEEmembership{Member,~IEEE,} Abdullah Gani,~\IEEEmembership{Senior Member,~IEEE}, Rajkumar Buyya,~\IEEEmembership{Senior Member,~IEEE}

\thanks{Manuscript received Dec 18, 2012; revised March 05, 2013 and 06 May, 2013;This work is funded by the Malaysian Ministry of Higher Education under the University of Malaya High Impact Research Grant - UM.C/HIR/MOHE/FCSIT/03. Ejaz Ahmed's research work is supported by the Bright Spark Unit, University of Malaya, Malaysia.} 

\thanks{Saeid Abolfazli(corresponding author), Zohreh Sanaei, Ejaz Ahmed, and Abdullah Gani are with the Department of Computer System \& Technology, The University of Malaya, Kuala Lumpur, Malaysia (e-mail: \{abolfazli,sanaei, ejazahmed\}@ieee.org; abdullah@um.edu.my)}

\thanks{RajKumar Buyya is with the Department of Computing and Information Systems, The University of Melbourne, 111, Barry Street, Carlton, Melbourne, VIC 3053, Australia, Email: raj@csse.unimelb.edu.au}
}
\markboth{IEEE Communications Surveys \& Tutorials, ACCEPTED FOR PUBLICATION}
{Shell \MakeLowercase{\textit{et al.}}: Bare Demo of IEEEtran.cls for Journals}
\maketitle


\begin{abstract}
Recently, Cloud-based Mobile Augmentation (CMA) approaches have gained remarkable ground from academia and industry. CMA is the state-of-the-art mobile augmentation model that employs resource-rich clouds to increase, enhance, and optimize computing capabilities of mobile devices aiming at execution of resource-intensive mobile applications. Augmented mobile devices envision to perform extensive computations and to store big data beyond their intrinsic capabilities with least footprint and vulnerability. Researchers utilize varied cloud-based computing resources (e.g., distant clouds and nearby mobile nodes) to meet various computing requirements of mobile users. However, employing cloud-based computing resources is not a straightforward panacea. Comprehending critical factors (e.g., current state of mobile client and remote resources) that impact on augmentation process and optimum selection of cloud-based resource types are some challenges that hinder CMA adaptability. This paper comprehensively surveys the mobile augmentation domain and presents taxonomy of CMA approaches. The objectives of this study is to highlight the effects of remote resources on the quality and reliability of augmentation processes and discuss the challenges and opportunities of employing varied cloud-based resources in augmenting mobile devices. We present augmentation definition, motivation, and taxonomy of augmentation types, including traditional and cloud-based. We critically analyze the state-of-the-art CMA approaches and classify them into four groups of distant fixed, proximate fixed, proximate mobile, and hybrid to present a taxonomy. Vital decision making and performance limitation factors that influence on the adoption of CMA approaches are introduced and an exemplary decision making flowchart for future CMA approaches are presented. Impacts of CMA approaches on mobile computing is discussed and open challenges are presented as the future research directions.
\end{abstract}


\section{Introduction}
\IEEEPARstart{S}{ince} a decade ago, the visions of \textit{`information under fingertip'} \cite{Weiser2002} and \textit{`unrestricted mobile computing'} \cite{Satyanarayanan1997} aroused the need to enhance computing power of mobile devices to meet the insatiable computing demands of mobile users \cite{SaeidAbolfazli2012}. 
In the late 90s, the concept of load sharing and remote execution aimed to augment computing capabilities of mobile devices by shifting the resource-intensive mobile codes to surrogates (powerful computing device(s) in vicinity) \cite{Othman:1998:PCS:584007.584011, Rudenko1998a,Satyanarayanan2001}. Although remote execution efforts \cite{Flinn, Flinnc,Balan2006,Balan2002cyber,Balan, Balana, Goyal2004a,Kristensen,Kristensen2008, Su2005,Gu2003, Kristensen2010} have yielded many impressive achievements, several challenges such as reliability, security, and elasticity of surrogates hinder the remote execution adaptability \cite{Sharifi2011}. For instance, the resource sharing and computing services of surrogates can be terminated without prior notice and their content can be accessed and altered by the surrogate machine or other users in the absence of a Service Level Agreement (SLA). SLA is a formal contract employed and negotiated in advance between service provider and consumer to enforce certain level of quality against a fee. 

\begin{figure*}[t]
\centering
\includegraphics[scale=0.62]{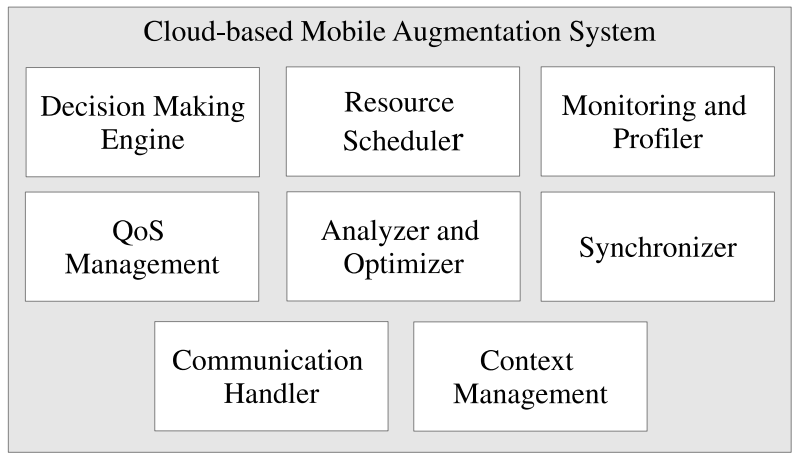}
\caption{Major Building Blocks of an Exemplary CMA System.} \label{CMA building blocks}
\end{figure*}

Few years later, emergence of cloud resources created an opportunity to mitigate the shortcomings of utilizing surrogates in augmenting mobile devices. Cloud is a type of distributed system comprised of a cluster of powerful computers accessible as unified computing resource(s) based on an SLA \cite{buyya2009cloud}. Cloud computing as ``a model for enabling ubiquitous, convenient, on-demand network access to a shared pool of configurable computing resources (e.g., networks, servers, storage, applications, and services) that can be rapidly provisioned and released with minimal management effort or service or service provider interaction" \cite{Mell2011} stimulates researchers to adopt the cutting edge technology in mobile device augmentation: Cloud-based Mobile Augmentation (CMA). Cloud-based Mobile Augmentation (CMA) is the-state-of-the-art mobile augmentation model that leverages cloud computing technologies and principles to increase, enhance, and optimize computing capabilities of mobile devices by executing resource-intensive mobile application components in the resource-rich cloud-based resources. Cloud-based resources include varied types of mobile/immobile computing devices that follow cloud computing principles \cite{Buyya2010, HoganJuly2011} to perform computations on behalf of the resource-constraint mobile devices. Figure \ref{CMA building blocks} depicts major building blocks of a typical CMA system. It is notable that these building blocks are optional superset, and specific CMA system may not have all these building blocks.

CMA efforts \cite{Huerta-Canepa, cuervo2010maui,Satyanarayanan2009, Verbelen2012,Hung2012,Kempa, Kosta2011, Guo2011, A.Manjunatha2010,Zhang2011, Chun2011, Chun2009,March2011,Luo2009, Badidi2011, Liu2009, Kumar2010,Chuna,Lu2011,Kemp2010a, Kemp2010, MOMCC, SAMI, Ma2012, Verbelen2012,Gu2012,Giurgiu2009} exploit various cloud-based computing resources, especially distant clouds and proximate mobile nodes to augment mobile devices. Distant clouds are giant clouds such as Amazon EC2\footnote{http://aws.amazon.com/ec2/} located inside the vendor premise \textemdash far away from the mobile clients\textemdash offering infinite, elastic computing resources with extreme computing power and high WAN (Wide Area Network) latency. Proximate mobile nodes are building a cluster of mobile computing devices scattered near the mobile clients offer limited computing power with lower WAN latency than distant clouds. 

\begin{table}[t] 
\caption{List of Acronyms and Corresponding Full Forms.}\label{acronyms}
\small
  \centering
    \begin{tabular}{|l| l|}
    \hline
\multicolumn{1}{|c|}{\textbf{Acronym}}     &  \multicolumn{1}{|c|}{\textbf{Full form}} \bigstrut\\
 \hline
     2D & 2 Dimensional \bigstrut\\
    \hline
     2G   & 2nd Generation   \bigstrut\\
    \hline
     3D   & 3 Dimensional   \bigstrut\\
    \hline
     3G   & 3rd Generation   \bigstrut\\
    \hline
      API & Application Programming Interface \bigstrut\\
    \hline
      App & Application (mobile application) \bigstrut\\
        \hline
     ARM   & Advanced RISC Machines   \bigstrut\\
    \hline
     CMA   & Cloud-based Mobile Augmentation   \bigstrut\\
    \hline
     CPU   & Central Processing Unit   \bigstrut\\
    \hline
     DSL   & Domain Specific Language   \bigstrut\\
    \hline
     DVMS   & Dynamic VM Synthesis   \bigstrut\\
    \hline
     FTP   & File Transfer Protocol   \bigstrut\\
    \hline
      GPU & Graphics Processing Unit \bigstrut\\
    \hline
     GUI   & Graphical User Interface   \bigstrut\\
    \hline
     I/O   & Input/Output   \bigstrut\\
    \hline
     IaaS   & Infrastructure as a Service \bigstrut\\
    \hline
     IP   & Internet Protocol   \bigstrut\\
    \hline
     IP TV   & Internet Protocol Television   \bigstrut\\
    \hline
     iSCSI   & Internet Small Computer System Interface \bigstrut\\
    \hline
     MCC   & Mobile Cloud Computing \bigstrut\\
    \hline
     MNO   & Mobile Network Operator   \bigstrut\\
    \hline
     OS   & Operating System   \bigstrut\\
    \hline
     P2P   & Peer-to-Peer  \bigstrut\\
    \hline
     PC   & Personal Computer   \bigstrut\\
    \hline
     QoS   & Quality of Service   \bigstrut\\
    \hline
     R\&D   & Research and Development   \bigstrut\\
    \hline
     RAM   & Random Access Memory   \bigstrut\\
    \hline
     RISC   & Reduced Instruction Set Computing   \bigstrut\\
    \hline
     RPC   & Remote Procedure Call \bigstrut\\
    \hline
     SAL   & Service Abstraction Layer   \bigstrut\\
    \hline
     SLA   & Service Level Agreement   \bigstrut\\
    \hline
     TCP   & Transmission Control Protocol   \bigstrut\\
    \hline
     UDDI & Universal Description Discovery and Integration   \bigstrut\\
    \hline
     UI   & User Interface   \bigstrut\\
    \hline
     VM   & Virtual Machine   \bigstrut\\
    \hline
     WAN   & Wide Area Network   \bigstrut\\
    \hline
     Wi-Fi   & Wireless Fidelity   \bigstrut[t]\\
     \hline
    \end{tabular}%
  \label{tab:addlabel}%
\end{table}%

Although heterogeneity among cloud-based resources increases service flexibility and enhances users' computing experience, determining the most appropriate computing resources among available options and performing upfront analysis of influential factors (e.g., user preferences and available native mobile resources) are critical in the adaptability of CMA approaches. Thus, `resource scheduler' and `analyzer and optimizer' components depicted in Figure \ref{CMA building blocks} are needed to analyze and allocate appropriate resources to each task in a typical CMA system. Moreover, several questions need to be addressed before the CMA concept can be successfully employed in the real scenarios. For instance, can CMA augment computing capabilities of mobile devices and save local resources to enhance user experience? Is CMA always feasible and beneficial? Which type of resources is appropriate for a certain task? Answering these questions requires `monitoring and profiler', `QoS management', `context management', and `decision making engine' components  to perform in each CMA system (see Figure \ref{CMA building blocks}). Therefore, an augmentation decision engine similar to those used in \cite{cuervo2010maui,Zhang2011,Giurgiu2009} and exemplary decision making flow presented in this paper (discussed in part \ref{feasibility}) to determine the mobile augmentation feasibility is needed to amend the CMA performance and reliability. During augmentation process, the local and native application state stack needs synchronization to ensure integrity between native and remote data. Upon successful outsourcing, remote results need to be returned and integrated to the source mobile device. Thus, the `Synchronizer'  component needs to perform in typical CMA approaches (see Figure \ref{CMA building blocks}).

Although CMA approaches can empower mobile processing and storage capabilities, several disadvantages such as application development complexity and unauthorized access to remote data demand a systematized plenary solution. Performance of the CMA systems is highly influenced by various challenges and issues of wireless networking and cloud computing technologies. CMA researchers require a high performance, elastic, robust, reliable, and foreseeable communication throughput between mobile nodes and cloud servers which is not yet realized despite of remarkable efforts and achievements of communication and networking societies. Current shortcomings and deficiencies of wireless communication and networking, especially seamless connectivity and mobility, high performance communication throughput provisioning, and wireless data interception discourage system analysts, engineers, developers, and entrepreneurs from deploying CMA-enabled mobile applications due to the high risk of system malfunction and user experience degradation.

Moreover, CMA systems require accurate estimation mechanisms to predict the overall time and energy consumption of communication and computation tasks while exploiting clouds. Such estimation is a challenging task considering huge infrastructures' performance diversity \cite{Ou2012} and policy heterogeneity \cite{Z2012} of cloud services in intermittent wireless environment. Despite of blooming efforts endeavoring to analyze and comprehend the cloud computing model and behavior \cite{Salah2012, Smith2012, DiFrancesco2012,Heide2012}, CMA solutions are still unable to accurately foresee required time and energy of exploiting cloud resources to execute intensive applications. Additionally, sundry cloud challenges, especially live VM migration, infrastructure and platform heterogeneity, efficient allocation of clouds to tasks, QoS management, security, privacy, and trust in cloud increase system complexity and decrease successful CMA systems adoption. 

Among limited studies of the domain, \cite{Sharifi2011} and \cite{Shiraz2012b} survey remote execution and application offloading algorithms with focus on how task offloading is performed in various efforts. Fernando et al. \cite{fernando2012mobile} and Dinh et al. \cite{Dinh2011} sought to explain the convergence of mobile and cloud computing, and distinguish it from the earlier domains such as cloud and grid computing \cite{Foster2001}. The authors describe issues, particularly mobile application offloading, privacy and security, context awareness, and data management. Sanaei, Abolfazli, Gani, and Buyya \cite{Z2012} present a comprehensive survey on MCC with major focus on heterogeneity. The authors describe the challenges and opportunities imposed by heterogeneity and identify hardware, platform, feature, API, and network as the roots of MCC heterogeneity. They explain major heterogeneity handling approaches, particularly virtualization, service oriented architecture, and semantic technology. However, the computing performance, distance, elasticity, availability, reliability, and multi-tenancy of remote resources are marginally discussed in these studies that necessitate further research to explain the impact of remote resources on augmentation process and highlight paradigm shift from the unreliable surrogates to reliable clouds. 

In this paper, we survey the state-of-the-art mobile augmentation efforts that employ cloud computing infrastructures to enhance computing capabilities of resource-constraint mobile devices, especially smartphones. To the best of our knowledge, this is the first effort that studies the impacts of cloud-based computing resources on mobile augmentation process. We differentiate augmentation from similar concepts of load sharing and remote execution, and present augmentation motivation. We review efforts that endeavor to mitigate the mobile devices' shortcomings and classify them as hardware and software to devise a taxonomy. The impacts of CMA in mobile computing are presented. The characteristics of cloud-based remote resources and their role in CMA effectiveness are studied and classified into four groups, namely distant immobile clouds, proximate immobile computing entities, proximate mobile computing entities, and hybrid based on their mobility and physical location traits. Furthermore, the state-of-the-art CMA models are reviewed and taxonomized into four classes of distant fixed, proximate fixed, proximate mobile, and hybrid according to our cloud-based resource classification. Factors impact on the CMA adaptability are identified and described as augmentation environment, user preferences and requirements, mobile devices, cloud servers, and contents. Five major metrics that limit the performance of CMA approaches are studied. A sample flowchart of decision making engines for imminent CMA solutions is presented and several open challenges are discussed as the future research directions. Such survey is beneficial to the communication and networking communities, because comprehending CMA process and current deployment challenges are beneficial in modifying the fundamental networking infrastructures to optimize the CMA process. In this paper, we use the terms mobile devices and smartphones interchangeably with similar notion. Table \ref{acronyms} shows the list of acronyms used in the paper.

The remainder of this paper is organized as follows. Section \ref{mobileaugmentation} introduces mobile computation augmentation, presents its motivation and describes the taxonomy of mobile augmentation types. The impacts of CMA on mobile computing are presented in Section \ref{impactofcma}. Section \ref{cloud-based resources} presents the analysis and taxonomy of varied cloud-based augmentation resources. Comprehensive survey of the state-of-the-art CMA approaches is presented and taxonomy is devised in Section \ref{state-of-the-art}. We discuss the CMA decision making and limitation factors and illustrate CMA feasibility in Section \ref{sec:cma-prospectives}. Finally, open research challenges are presented in Section \ref{openissues} and paper is concluded in Section \ref{conclusions}. 

\section{Mobile Computation Augmentation} \label{mobileaugmentation}
In this Section, we present a definition on mobile computing augmentation based on the available definitions on the relevant concepts, particularly remote execution \cite{Rudenko1998a} and cyber foraging \cite{Satyanarayanan2001}. Additionally, the motivation for performing mobile computation augmentation is described and taxonomy of mobile augmentation types is presented.

\subsection{Definition} \label{definition}
Indeed, empowering computation capabilities of mobile devices is not a new concept and there have been different approaches to achieve this goal, including load sharing \cite{Othman:1998:PCS:584007.584011}, remote execution \cite{Rudenko1998a}, cyber foraging \cite{Satyanarayanan2001}, and computation offloading \cite{Li2001, Li2002} that are described as follows. {We have analyzed them and summarized the analysis results in Table \ref{approaches}. Results in this Table are extracted from the early efforts in each category, which are already deviated from their original characteristics due to the research achievements.}

\begin{table*}[t] 
\caption{Initial Features of Mobile Empowerment Approaches.} \label{approaches}
  \centering
    \begin{tabular}{|l|c|c|c|c|c|c|} 
    \hline
    \multicolumn{1}{|c|}{Approach} & \multicolumn{1}{c|}{Architecture} & \multicolumn{1}{c|}{Client Load} & \multicolumn{1}{c|}{Migration} & \multicolumn{1}{c|}{Partitioning} & \multicolumn{1}{c|}{Server} & \multicolumn{1}{c|}{Mobility}\bigstrut\\
    \hline
Load Sharing & \multicolumn{1}{c|}{Client-Server} & \multicolumn{1}{c|}{Entire Task} & \multicolumn{1}{c|}{Entire task} & \multicolumn{1}{c|}{NA} & \multicolumn{1}{c|}{Server} & \multicolumn{1}{c|}{NA}\bigstrut\\
    \hline
Remote  & \multicolumn{1}{c|}{\multirow{2}[2]{*}{Client-Server}} & \multicolumn{1}{c|}{\multirow{2}[2]{*}{Entire Task}} & \multicolumn{1}{c|}{\multirow{2}[2]{*}{Entire/partial}} & \multicolumn{1}{c|}{\multirow{2}[2]{*}{Static}} & \multicolumn{1}{c|}{Server} & \multicolumn{1}{c|}{No}\bigstrut[t]\\
Execution & \multicolumn{1}{c|}{} & \multicolumn{1}{c|}{} & \multicolumn{1}{c|}{} & \multicolumn{1}{c|}{} & \multicolumn{1}{c|}{/desktop} & \multicolumn{1}{c|}{}\bigstrut[b]\\
    \hline
   Cyber & \multicolumn{1}{c|}{Client-Server} & \multicolumn{1}{c|}{\multirow{2}[2]{*}{Entire Task}} & \multicolumn{1}{c|}{\multirow{2}[2]{*}{Entire/partial}} & \multicolumn{1}{c|}{\multirow{2}[2]{*}{Dynamic}} & \multicolumn{1}{c|}{Surrogates} & \multicolumn{1}{c|}{No}\bigstrut[t]\\
Foraging& \multicolumn{1}{c|}{Peer-to-Peer} & \multicolumn{1}{c|}{} & \multicolumn{1}{c|}{} & \multicolumn{1}{c|}{} & \multicolumn{1}{c|}{} & \multicolumn{1}{c|}{}\bigstrut[b]\\
    \hline
   Mobile & \multicolumn{1}{c|}{Varies, e.g.,} & \multicolumn{1}{c|}{\multirow{5}[2]{*}{/Nil}} & Entire/partial/ & \multicolumn{1}{c|}{Static \&} & \multicolumn{1}{c|}{Server,} & \multicolumn{1}{c|}{Yes}\bigstrut[t]\\
    Computation  & \multicolumn{1}{c|}{Client-server}& \multicolumn{1}{c|}{Entire/partial} & Nil migration & \multicolumn{1}{c|}{dynamic} & \multicolumn{1}{c|}{surrogate}& \multicolumn{1}{c|}{} \\
    Augmentation & \multicolumn{1}{c|}{P2P, Adhoc} & \multicolumn{1}{c|}{} &(Use remote&    & \multicolumn{1}{c|}{\&mobile} & \multicolumn{1}{c|}{} \\
       & \multicolumn{1}{c|}{collaborative} & \multicolumn{1}{c|}{} & services) &    &  & \multicolumn{1}{c|}{}\\ \hline
    \end{tabular}%
\end{table*}%

\textbullet \textbf{ Load Sharing:} Othman and Hailes' work \cite{Othman:1998:PCS:584007.584011} in 1998 can be considered as one of the earliest efforts to conserve native resources of mobile devices using a software approach. The main idea is inspired from the concept of load balancing in distributed computing which is ``a strategy which attempts to share loads in a distributed system without attempting to equalize its load'' \cite{Othman:1998:PCS:584007.584011}. This approach migrates the whole computation job for remote execution. It considers several metrics such as job size, available bandwidth, and result size to identify if the load balancing and transferring the job to the remote computer can save energy. However, they need to send the task and data to the nearest base station and wait for the results to return. The base station is responsible to find appropriate server to run the job and forward the results back to the mobile device. Moreover, computing server is a fixed computer and there is no provision for user and code mobility at run time.

\textbullet \textbf{ Remote Execution:} The concept of remote execution for mobile clients emerged in 90’s and several researchers \cite{ Athan1993, Rudenko1998a, Bakre1995, Rudenko1999, Wrabetz1995} endeavor to enable mobile computers performing remote computation and data storage to conserve their scarce native resources and battery. In 1998 \cite{Rudenko1998a}, feasibility of the remote execution concept on mobile computers, particularly laptops was investigated. The authors report that remote execution can save energy if the remote processing cost is lower than local execution. Remote execution involves migrating computing tasks from the mobile device to the server prior the execution. The server performs the task and sends back the results to the mobile device. However, difference between computation power of client and server is not a metric of decision making in this method. Moreover, the whole task needs to be migrated to the remote server prior the execution which is an expensive effort. It also neglects the impact of environment characteristics on the remote execution outcome. Static decision making is another shortcoming of this proposal.

\textbullet \textbf{ Cyber Foraging:} Satyanarayana in 2001 \cite{Satyanarayanan2001} further developed the remote execution idea by considering dynamism in remote execution process. The author defined cyber foraging as the process ``to dynamically augment the computing resources of a wireless mobile computer by exploiting wired hardware infrastructure''. Resources in cyber foraging are stationary computers or servers in public places \textemdash connected to wired Internet and power cable\textemdash that are willing to perform intensive computation on behalf of the resource-constraint mobile devices in vicinity. 

However, load sharing, remote execution, and cyber foraging approaches assume that the whole computing task is stored in the device and hence, it requires the intensive code and data to be identified and partitioned for offloading \textemdash either statically prior the execution or dynamically at runtime \textemdash which impose large overhead on resource-poor mobile device \cite{Sharifi2011}. Moreover, as Kumar et al. \cite{Kumar2012} explain, for each mobile user that runs the intensive application, the whole offloading process must be repeated including decision making process in the device and transferring the heavy components and large data to the network. Due to slight differences among these concepts, researchers use the terms `remote execution', `cyber foraging', and `computation offloading' interchangeably in the literature with similar principle and notion.

Nevertheless, researchers in recent activities  \cite{Lu2011,March2011, MOMCC, SAMI} aim to enhance computing capabilities of mobile devices in a slightly different manner. They assume to store the intensive code and data outside the device and keep the rest in the mobile device instead of storing the whole task \textemdash including both lightweight and intensive code and data \textemdash in the mobile device. Therefore, the overhead of identifying, partitioning, and migrating the resource-intensive task is mitigated, energy is saved, and storage problem is alleviated in mobile devices. Moreover, storing intensive components outside the device, in a publicly accessible storage, can increase their reusability and enable more than one user to leverage their computation services. Therefore, we coin the term \textit{mobile computation augmentation} as the wider phrase that subsumes load sharing, remote execution, cyber foraging, and other approaches that augment computing capabilities of mobile devices. 

\textbullet \textbf{ Mobile Computation Augmentation:}
Mobile computation augmentation, or augmentation in brief, is the process of increasing, enhancing, and optimizing computing capabilities of mobile devices by leveraging varied feasible approaches, hardware and software. Mobile device is any non-stationary, battery-operating computing entity able to interact with end-user and execute transactions, store data, and communicate with the environment using wireless technologies and varied sensors. Smartphone, Tablet, handheld/wearable computing devices, and vehicle mount computers are mobile device instances. Approaches that can augment mobile devices include hardware and software. Hardware approach involves manufacturing high-end physical components, particularly CPU, memory, storage, and battery. Software approaches can be \textemdash but are not limited to \textemdash computation offloading, remote data storage, wireless communication, resource-aware computing, fidelity adaptation, and remote service request (e.g., context acquisition). 

Augmentation approaches can increase computing capabilities of mobile devices and conserve energy. They can be leveraged in three main categories of applications, namely (i) computing-intensive software such as speech recognition and natural language processing, (ii) data-intensive programs such as enterprise applications, and (iii) communication-intensive applications such as online video streaming applications. The augmented mobile device is able to perform complex tasks that could not otherwise perform. Hence, the mobile application developers do not take into account resource shortcomings of mobile devices in developing mobile application and users will not consider their devices weaknesses in utilizing varied complex applications. 

\begin{figure*}[t]
\begin{center}
\includegraphics[scale=0.2]{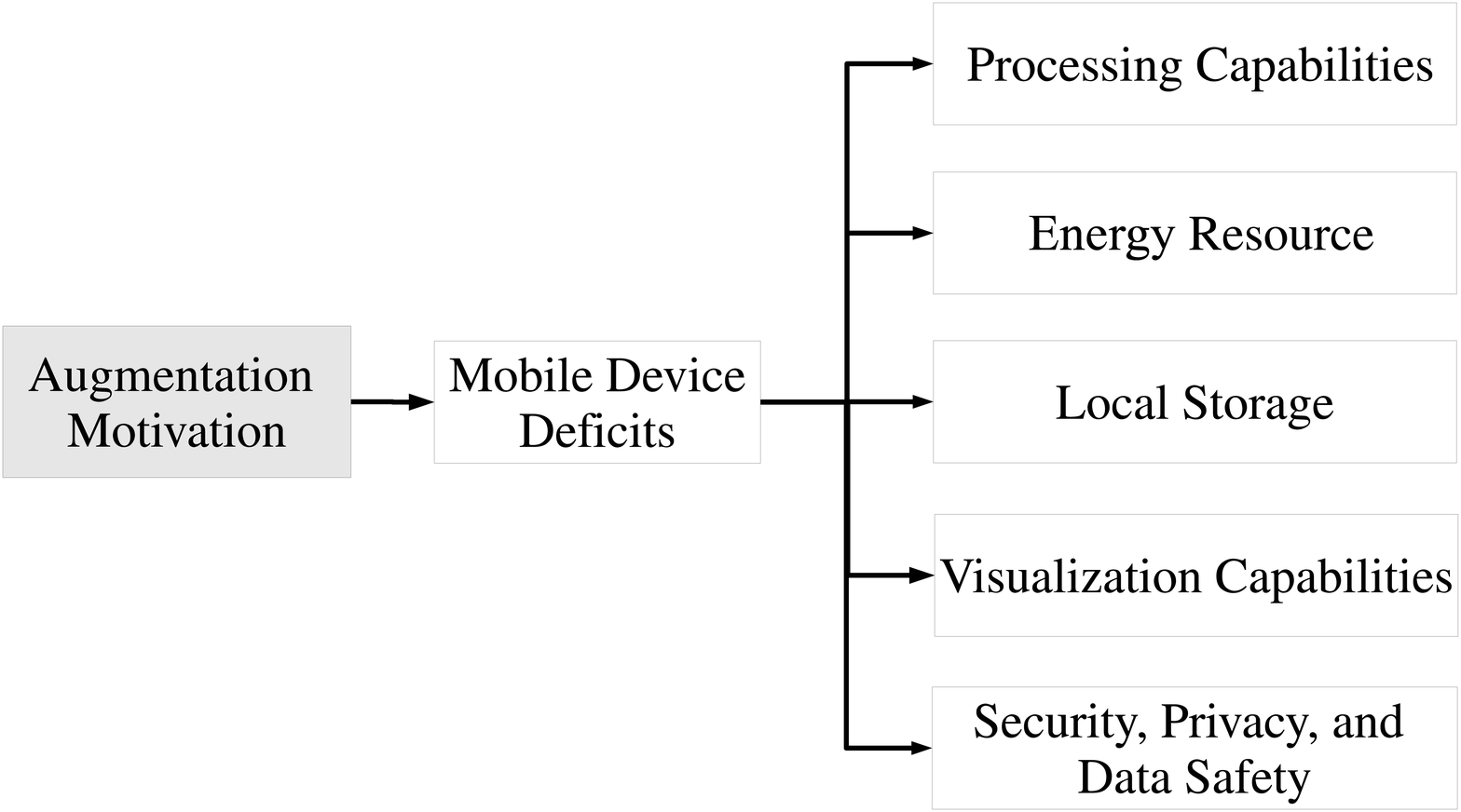}
\caption{Taxonomy of Augmentation Motivation: Intrinsic and non-intrinsic mobile challenges motivate augmentation.} \label{augmentation motivation}
\end{center}
\end{figure*}

\subsection{Motivation} \label{motivation}
Mobile devices have recently gained momentous ground in several communities like governmental agencies, enterprises, social service providers (e.g., insurance, Police, fire departments), healthcare, education, and engineering organizations \cite{Bent2012, Khalifa2011}. However, despite of significant improvement in mobiles' computing capabilities, still computing requirements of mobile users, especially enterprise users, is not achieved. 

Several intrinsic deficiencies of mobile devices encumber feasibility of intense mobile computing and motivate augmentation. Leveraging augmentation approaches, vision of performing intense mobile operations and control such as remote surgery, on-site engineering, and visionary scenarios similar to the lost child and disaster relief described in \cite{Satyanarayananb} will become reality. In this part, we analyze and taxonomize smartphones' deficits that can be alleviated by augmentation. Figure \ref{augmentation motivation} depicts our devised taxonomy.

\subsubsection{Processing Power}
Processing deficiencies of mobile clients due to slow processing speed and limited RAM is one of the major challenges in mobile computing \cite{Satyanarayananb}. Mobile devices are expected to have high processing capabilities similar to computing capabilities of desktop machines for performing computing-intensive tasks to enrich user experience. Realizing such vision requires powerful processor being able to perform large number of transactions in a short course of time. 

Large internal memory/RAM to store state stack of all running applications and background services is also lacking. Beside local memory limitations, memory leakage also intensifies memory restrains of mobile devices. Memory leakage is the state of memory cells being unnecessarily occupied by running applications and services or those cells that are not released after utilization. Garbage-collector-based languages like Java in Android\footnote{url{http://www.android.com/}} contribute to memory leakage due to failed or delayed removal of unused objects from the memory \cite{Park2012}. Android's kernel level transactions can also leak memory in the absence of memory management mechanisms \cite{Xu2008,Park2012}. Moreover, inward deficiency and inefficient design and implementation of mobile applications can also waste scarce memory of mobile devices. Thus, in the absence of required memory, applications are frequently paused or terminated by the operating system leading to longer execution time, excessive resource dissipation, and ultimately mobile user experience degradation.

\subsubsection{Energy Resources}
Energy is the only non-replenishable resource in mobile devices that demands external resources to be replenished \cite{Satyanarayanan2005, Miettinen2010}. Currently, energy requirement of a mobile device is supplied via lithium-ion battery that lasts only few hours if device is computationally engaged. Battery capacity is increasing at about 5 to 10\% a year \cite{Neuvo2004, Robinson2009} as battery cells are excessively dense \cite{Satyanarayanan2005}. Moreover, mobile device manufacturers endeavor to attain device lightness, compactness, and handiness, which prevent exploitation of bulky long-lasting batteries. User safety is another concern that confines manufacturers to produce low capacity batteries \cite{mobileexplosion}. While explosion of a battery with few hundreds milliamperes capacity can jeopardize human life \cite{cellularnews}, explosion of a high-capacity battery can carry catastrophic consequences. 

Energy harvesting efforts \cite{Flinna, Starner, R.Avro2009} seek to replenish energy from renewable resources, particularly human movement, solar energy, and wireless radiation, but these resources are mostly intermittent and not available on-demand \cite{Pickard2012}. For instance, a sitting mobile user at night cannot have any external power source in the absence of wall power and wireless radiations. Moreover, researchers aim at reducing the energy overhead in different aspects of computing, including hardware, OS, application, and networking interface \cite{Bianzino2012,Vallina-Rodriguez2013}. Efforts are directed to develop alternative energy resources such as nuclear batteries that will likely last months or years \cite{KelseyJackson2009}. However, significant deal of R\&D is needed to fulfill ever-increasing energy requirements of mobile users.

Hence, in the absence of long-spanning energy on mobile devices, alternative augmentation approaches play a vital role in maturing mobile and ubiquitous computing.
 
\subsubsection{Local Storage}
Drastic increasing in the number of applications and amount of digital contents such as pictures, songs, movies, and home films \cite{Gantz2008} from one hand and limited storage of mobile devices from the other hand decelerate usability of mobile devices. While PCs are able to locally store huge amount of data, smartphones are limited to few gigabytes of space which are mostly occupied by system files, user applications, and personal data. Therefore, frequent storing, updating, and deleting data as well as uninstalling and reinstalling applications due to space limitation cause irksome impediments to mobile users \cite{Guo2008}. Additionally, delivering offline usability, which is one of the most important characteristics of contemporary applications, remains an open challenge since mobile devices lack large local storage. 

\subsubsection{Visualization Capabilities} \label{visualization} 
Effective data visualization on small mobile devices' screen is a non-trivial task when current screen manufacturing technologies and energy limitations do not allow significant size extensions without losing device handiness. Currently smartphones like HTC One X\footnote{http://www.htc.com/www/smartphones/htc-one-x/} and Samsung Galaxy Note II\footnote{http://www.samsung.com/my/consumer/mobile-devices/galaxy-note/galaxy-note/GT-N7100RWDXME} have the biggest screens, at 4.7 and 5.5 inches respectively; however, they are very small compared to PCs and notebooks. 

Therefore, efficient data visualization in small smartphones' screen necessitates software-based techniques similar to tabular pages, 3D objects, multiple desktops, switching between landscape and portrait views (needs accelerometer), and verbal communication to virtually increase presentation area. Recently, computing-intensive mobile 3D display technology is promising to noticeably mitigate the visualization deficit of contemporary smartphones. Glass-free auto-stereoscopic displays \cite{Ortiz2011} can present 3D data by exploiting binocular parallax to offer a different view for each eye. Taking advantages of current and imminent software-based techniques beside native tools, especially tilting sensors significantly improve the mobile visualization capabilities in the near future. However, these approaches are computation-intensive processes that quickly drain battery \cite{Ortiz2011,Capin2008}. A feasible alternative solution to realize software-based content presentation approaches is to augment smartphones' computing capabilities.

\subsubsection{Security, Privacy, and Data Safety}
Mobile end-users are concerned about security and privacy of their personal data, banking records, and online behaviors \cite{Survey2011}. The dramatic increase in cybercrime and security threats within mobile devices \cite{LaPollatobepublished}, cloud resources \cite{Xiao2013} and wireless transactions makes security and privacy more challenging than ever \cite{CHRISTIANCACHIN2011}. Moreover, performing complex cryptographic algorithms is likely infeasible because of computing deficiencies of mobile devices. Securing files using pair of credentials is also less realistic in the absence of large keyboard.

Data safety is another challenge of mobile devices, because information stored inside the local storage of mobile devices are susceptible to safety breaches due to high probability of hardware malfunction, physical damage, stealing, and loss. 
\\
\\
\noindent Amalgam of these problems and deficiencies in mobile computing stimulates researchers from academia and industry to exploit novel technologies and approaches to augment computing capabilities of mobile devices which is subject of this study.

\begin{figure*} [t]
\centering
\includegraphics[scale=0.18]{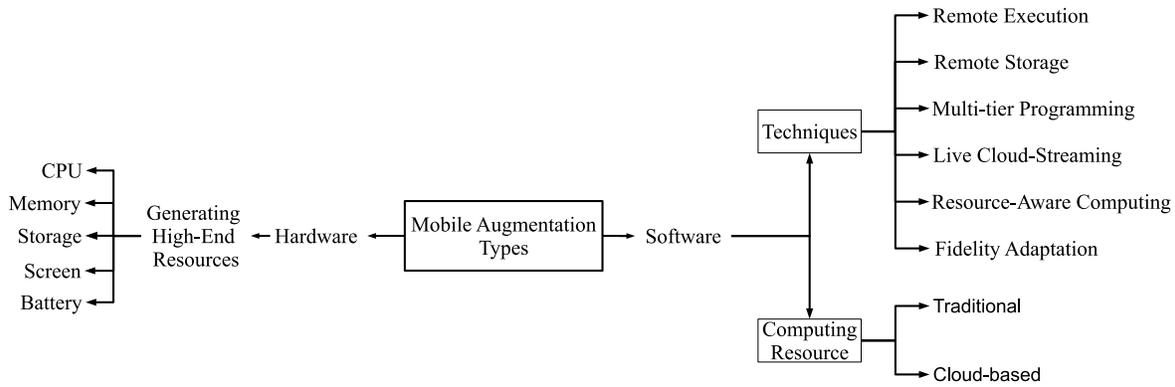}
\caption{Taxonomy of Mobile Augmentation Types.} \label{mobile augmentation}
\end{figure*}

\subsection{Mobile Augmentation Types: Taxonomy} \label{taxonomyofaugmentationapproach}
In this Section, we analyze and classify augmentation approaches into two major types of hardware and software. Our devised taxonomy is depicted in Figure \ref{mobile augmentation} and described as follows. 

\noindent \textbf{Hardware.} \label{hardware augmentation}
The hardware approach aims to empower smartphones by exploiting powerful resources, particularly multi-core CPUs with high clock speed \cite{multicore-nvidia}, large screen, and long-lasting battery \cite{ARM2009,KelseyJackson2009}.  
ARM\footnote{\url{http://arm.com}} and Samsung\footnote{\url{http://samsung.com}} are major mobile processor manufacturers producing multi-core processors such as ARM Cortex-59\footnote{\url{http://www.arm.com/products/processors/cortex-a50/cortex-a57-processor.php}} and Samsung Exynos 5 Octa core\footnote{\url{http://www.samsung.com/global/business/semiconductor/minisite/Exynos/blog\_CES\_2013\_Samsung\_Mobilizes\_Possibility\_with\_Exynos5Octa.html}} 
that perform in higher speed than a single core processors \cite{multicore-nvidia}. However, doubling the CPU clock speed approximately octuples the device energy consumption \cite{Kumar2012}. 

Nevertheless, augmentation via sophisticated hardware is hindered by several obstacles. Firstly, generating powerful processor, large storage, and big screen decrease smartphone handiness due to additional heat, size, and weight. Secondly, considering the fact that utilizing long-lasting battery in small mobile devices is not feasible with current technologies, resource enlargement contributes toward faster battery drainage and shorter battery life time. Thirdly, equipping mobile devices with high-end hardware noticeably increases their price compare to the stationary machines. Unlike PCs, smartphone's hardware is not upgradable; hence, a new device should be possessed in case of technology advancement. Therefore, in the absence of futuristic engineering technologies, the hardware-based augmentation process is slow and expensive that necessitates alternative augmentation approaches to enhance computing capabilities of mobile devices without drastic ownership price hike.

\noindent \textbf{Software.}  \label{Softwareaugmentation}
Software-oriented mobile augmentation approaches are classified into five groups and will be explained later in this part. Resources that are used in major software-oriented approaches are classified into two groups, namely traditional and cloud-based. Their major differences lie on resource provisioning and access strategies, service security and delivery models, and resource characteristics. In traditional approaches, researchers leverage centralized resources of distant traditional servers or free nearby surrogates. Several problems such as resource availability, elasticity, and security of traditional approaches hinder their success. For instance, surrogates can terminate their services anytime without considering their current load, and can violate user security and privacy by changing execution sequence or altering raw and processed data.
 
To alleviate the problems of traditional servers, researchers in recent efforts \cite{Chuna, Chun2009, Kemp2010, Kemp2010a, Kempa,Guo2011, Zhang2011, cuervo2010maui,Chun2011,Verbelen2012,Altamimi2012} exploit highly available, elastic, secure cloud infrastructure. ``Cloud is a type of parallel and distributed system consisting of a collection of interconnected and virtualized computers dynamically provisioned and presented as one or more unified computing resources based on service-level agreements established through negotiation between the service provider and consumers" \cite{buyya2009cloud}.

While utilizing cloud resources, users pay for the amount and duration they utilize various resources (e.g., CPU, memory, and bandwidth) based on an agreed SLA. In the SLA, the amount and quality of required resources such as processor, RAM, and storage are specified and user is billed accordingly. Service delivery failure will be compensated by the vendor. Lucrative financial benefits of cloud services encourage cloud providers to compete in delivering high service availability, reliability, security, and robustness to increase their market share. Hence, the augmentation performance is less affected by resource unreliability and interruption.

Moreover, cloud infrastructures are available to end-users via Virtual Machine(VM)\footnote{http://www.vmware.com/virtualization/what-is-virtualization.html} to increase resource utilization ratio and enhance overall security and privacy. Virtualization technology aims to enable resource sharing in an isolated environment called VM. It realizes execution of multiple operating systems on a single machine and enables sharing of large resources among multiple end-users. Users can only access to infrastructures allocated to their VMs and cannot access prohibited resources and contents.

 \begin{table*} [!t]
\centering
\caption{Comparison Between Traditional and Cloud-Based Computing Resource.} \label{traditionalVScloudbased}
\begin{tabular}{|l|c|c|} 
\hline
Features & Traditional & Cloud-Based \\ \hline
Computation Power& Low & High \\ \hline
Elasticity & Low & High \\ \hline
User Experience & Low & High \\ \hline
Reliability & Low & High \\ \hline
Availability & Intermittent& On-demand \\ \hline
Client Mobility& Limited & Unlimited \\ \hline
Multi-tenancy & Not available  & Available \\ \hline
Serving Incentive & Not provisioned & Provisioned \\ \hline
Utilization Cost & Free & Pay-As-You-Use \\ \hline
Utilization Overhead & High & Low \\ \hline
Management & Decentralized & De/Centralized \\\hline
Back-end Connectivity & Wired & Wired \& Wireless\\ \hline
Communication Latency & Low & Varied \\ \hline
Computation Latency & High & Low \\ \hline
Security & Low & High \\  \hline
Data Safety & Low & High \\ \hline
\end{tabular}
\end{table*}

Table \ref{traditionalVScloudbased} summarizes the comparison results of traditional and cloud-based resources and advocates differentiations between the conventional servers and clouds. High computing power, elasticity, mobility support, low utilization overhead, and security are some of the significant advantages of cloud resources compare to the surrogates that advocate the latest paradigm shift in mobile augmentation.

Software augmentation techniques are classified as remote execution (offloading or cyber foraging) \cite{Satyanarayanan2001,Rudenko1998a,cuervo2010maui,Flinn, Balan2002cyber, Balana, Flinnc, Goyal2004a, Su2005, Balan, Gu2003, Chuna, Kempa, Kemp2010, Chun2009, Kemp2010a, Kristensen2010, Zhang2011, Chun2011, Kosta2011,Fekete2012}, remote storage \cite{Park2003}, Multi-tier programming\cite{March2011,SAMI,MOMCC}, live cloud-streaming \cite{Lawton2012}, resource-aware computing \cite{Seshasayee2007, Hong2009a}, and fidelity adaptation \cite{Krishna2004} and explained as follows. 

\textbullet\textit{ Remote execution:} As explained in \ref{definition},the resource-hungry components of mobile applications \textemdash in whole or part \textemdash are migrated to the resource-rich computing device(s) that are willing to share their resources with mobile devices. Rapid development of heterogeneous mobile devices obliges adaptive offloading approaches able to enhance capabilities of wide range of mobile devices in dynamic environment with least processing overhead and latency. The efficiency of offloading approaches highly depends on what component(s) can be partitioned? When partitioning takes place? Where to execute the component(s)? And how to communicate with the remote server? \cite{Murarasu2009}. Offloading approaches perform varied-time analysis to answer these questions, which are classified into three groups and explained as below.

\textit{Design Time Analysis:} In this method, the application's complexity is analyzed at design time to determine the answer of four above questions. Application developer or a middleware specifies the resource-intensive components of mobile application that can be offloaded to the remote server and label them as remote component(s). Programmers decide how to partition application and adapt its performance to the dynamic mobile environment which is a non-trivial task, mainly due to the lack of knowledge about the execution environment. Performing such action needs excessive programming skill and knowledge of computation offloading. Design time approaches \cite{Balana,Balan2002cyber,Flinnc} notably save native resources of mobile device by reducing the processing and monitoring overheads. However, partitioning prior to the execution is not always optimal and cannot accurately adapt performance in diverse execution environments and also imposes extra efforts on the application developer or middleware for deciding on partitioning. Hence, design time partitioning approaches are likely become obsolete.

\textit{Runtime Analysis:} Runtime or dynamic partitioning referred to methods such as \cite{cuervo2010maui,Abebe2012} that aims to answer four questions at runtime. They identify and partition the resource-hungry parts of the application, specify how and where to execute the partitioned components \cite{Salehie2009,Murarasu2009}, and determine how to communicate with the server. In dynamic methods, resource requirement of the application is analyzed and available resources are detected to decide if the application requires remote resources. Upon decision making the system performs offloading. Further monitoring is necessary to gather knowledge of available remote resources to maintain execution history. Although these approaches provide dynamic and flexible solutions, large amount of resources are wasted at runtime that prolongs application execution and decrease energy efficiency.

\textit{Hybrid Analysis:} The ultimate aim of hybrid approaches \cite{Huerta-Canepa2008} is to increase performance and efficiency of augmentation methods. Deciding on how to perform the offloading mainly depends on the native resources, remote resources, and available network bandwidth. In \cite{Huerta-Canepa2008}, prior to the application execution, the system decides based on four options, namely i) no action, ii) dynamic, iii) static, and iv) profile only whether to offload or not and in case of offloading specify what type of partitioning should take place. The profile only option is similar to the no action, but the systems collect execution information to maintain execution history for future purpose.

\textbullet\textit{ Remote Storage:} Remote storage is the process of expanding storage capability of mobile devices using remote storage resources. It enables maintaining applications and data outside the mobile devices and provides remote access to them. In early efforts, researchers in \cite{Park2003} utilize iSCSI (Internet Small Computer System Interface) \cite{Schmidt1997} \textemdash as a well-established protocol for remote storage \textemdash to access the server's I/O resources via mobile clients over the TCP/IP network to store, backup, and mirror data \cite{Lu2003}. However, the throughput of iSCSI is highly affected by the mobile-server distance. Using iSCSI is also difficult for handling large files such as multimedia and database files. Moreover, due to message passing in wireless medium through TCP/IP, the security and processing overhead (e.g., cryptography and data compression) are further challenges. To alleviate these challenges, several researches as MiSC \cite{Kim2005b}, UbiqStor \cite{Ok2006,Ok2005}, and Intermediate Target \cite{Kim2005} are proposed towards realizing remote storage on mobile devices. However, due to scalability, availability, performance, and efficiency issues of traditional servers, power of remote storage could not fully unleash using traditional servers. 

Several proposals and data storage services in academia and industry aim to expand mobile storage by exploiting cloud computing, especially Jupiter \cite{Guo2011}, SmartBox \cite{Zheng2010}, Amazon S3\footnote{\url{http://aws.amazon.com/s3/}}, Mozy\footnote{http://mozy.com}, Google Docs\footnote{https://docs.google.com/}, and DropBox\footnote{https://www.dropbox.com/}. For instance, Jupiter expands smartphone storage and assists end-users in organizing large applications and data. Jupiter leverages cloud infrastructures to store big data of mobile users. Heavy applications are executed inside the cloud's VM of smartphones and results are forwarded to the physical device after execution. Amazon S3 is a general purpose storage offering simple operations to store and retrieve cloud data while Mozy provides data backup facilities with main focus on enhancing cloud data safety against natural disasters. 

\textbullet\textit{ Resource-Aware Computing:} In resource-aware computing efforts, especially \cite{Kremer2003, Gurun2003, Ma2012b, Seshasayee2007, Hong2009a}, resource requirements of mobile applications are diminished utilizing the application-level resource management methods (using application management software such as compiler and OS) and lightweight protocols. Resource conservation is performed via efficient selection of available execution approaches and technologies \cite{Gurun2003}. Any mobile application consists of application-level resource management method is considered as a resource-aware application. For instance, in \cite{Hong2009a}, authors propose an energy-friendly scheme for content-based image retrieval applications using three offloading options, namely i) local extraction-remote search, ii) remote extraction-remote search, and iii) remote extraction-local search. The authors consider available bandwidth, image database size, and number of user queries to opt any of three offloading options for saving energy. In a high bandwidth network with limited queries, the third option is beneficial; the system uploads all un-indexed images  to the remote server and receives the results to be loaded into the memory. Then, all search queries are executed locally.

Similarly, applications can decide whether to choose 2G or 3G in telephony and FTP.  Using 2G network for telephony and 3G for FTP can noticeably reduce resource requirements of the mobile applications, according to the power consumption patterns presented in \cite{Perrucci}. 2G network technology consumes less energy for establishing a telephony communication, while 3G is more energy-friendly for file transfer transactions.

\textbullet\textit{ Fidelity adaptation:} Fidelity adaptation is an alternative solution to augment mobile devices in the absence of remote resources and online connectivity. In this method local resources are conserved by decreasing quality of application execution, which is unlikely desirable to end-users. As a well-known fidelity adaptation approach, we can refer to the YouTube\footnote{http://youtube.com}. Users in YouTube can adjust the streaming quality based on available bandwidth. To achieve optimized performance, researchers \cite{Flinna,Lara} leverage composition of cyber foraging and fidelity adaptation. 

\textbullet\textit{ Multi-tier Programming:} Developing distributed multi-tier mobile applications leveraging remote infrastructures is another technique employed in efforts such as \cite{SAMI, MOMCC, March2011,Christensen2009} to reduce resource requirements of mobile applications. The main idea in this type of mobile applications is to reduce the client-side computing workload and develop the applications with less native resource requirements. Certainly, the computationally intensive components of the applications are executed outside the device, whereas the interactive (user interface) and native codes (e.g., accessing to the device camera) remain inside the device for execution. 

Multi-tier applications are lightweight aiming to consume the least possible local resources by utilizing remote components and services, whereas native applications are monolithic applications often require runtime migration for execution. Therefore, monitoring time and communication overhead of multi-tier applications are shrunk leading to explicit resource saving and user experience enhancement. 

\textbullet\textit{ Live Cloud Streaming:} \label{cloud-streaming}
 In recent efforts to harness cloud resources, researchers from \textit{Onlive}\footnote{http://onlive.com} and \textit{Gaikai}\footnote{http://gaikai.com}, among other organizations introduce new approach to augment computing capabilities of mobile devices, entitled live cloud streaming \cite{Lawton2012}. In live cloud streaming approaches, mobile device acts as a dump client able to interact with server using a browser or application GUI. In live cloud streaming applications, entire processing take place in the cloud and results are streaming to the mobile devices. However, usability of cloud-streaming is hindered by latency, network bandwidth, portability, and network traffic cost. 

Functionality of cloud-streaming applications absolutely depends on the network availability and the Internet. Transferring mobile-user input to the server is another critical factor that requires considerable attention under wireless Internet connection. Moreover, since majority of mobile network providers deploy `pay-as-you-use' data plans, the large data traffic of cloud-streaming services imposes high communication cost on users. Yet congestion handling remains an open issue at peak hours. Entirely relying on cloud-streaming infrastructures and avoiding smartphones resources' utilization impact on application responsiveness and levy extravagant ownership, maintenance, power, and networking expenses to the cloud-streaming service providers, which is not a green computing approach.

\section{Impacts of CMA on Mobile Computing}   \label{impactofcma} 
This Section discusses the advantages and disadvantages of performing a CMA process on mobile computing that are summarized in Table \ref{Impacts}. We aim to demonstrate how CMA approaches mitigate deficiencies of mobile computing explained in Section \ref{motivation}. In this Section the terms `cloud resources' and `cloud infrastructures' refer to any type of cloud-based resources and infrastructures discussed in Section \ref{state-of-the-art}.

\subsection{Advantages}
In this part, eight major benefits of utilizing cloud resources in mobile augmentation processes are introduced. 

\subsubsection{Empowered Processing} \label{computingaugmentation}
Empowering processing is the state of virtually increased transaction execution per second and extended main memory leveraging CMA approaches. In computing-intensive mobile applications, either the hosting device does not have enough processing power and memory or cannot provide required energy. A common solution is to offload the application \textemdash in whole or part \textemdash to a reliable, powerful resource with least energy and time cost. In computation offloading, the complex, CPU- and memory-intensive components of a standalone application are migrated to the cloud. Consequently, the mobile devices can virtually perform and actually deliver the results of heavy transactions beyond their native capabilities. 

Although surrogates in traditional augmentation approaches \cite{Balana,Balan2002cyber,Balan2006,Flinnc} could increase computing capabilities of mobile devices, excessive overhead of arbitrary service interruption and denial could shadow augmentation benefits \cite{Sharifi2011,Shiraz2012c}. Cloud resources guarantee highest possible resource availability and reliability.
 
Leveraging CMA approaches, application developers build mobile application with no consideration on available native resources of mobile devices and mobile users dismiss their devices' inabilities. Hence, computing- and memory-intensive mobile applications like content-based image retrieval applications (enable mobile users to retrieve an image from the database) can be executed on smartphones without excess efforts. 

However, a flexible and generic CMA approach that can enhance plethora of mobile devices with least configuration, processing overhead, and latency is a vital need in excessively diverse mobile computing domain. Such diversity is mainly due to the rapid development of smartphones and Tablets, and sharp rise in their hardware, platform, API, feature, and network heterogeneity \cite{ZohrehSanaei2012} in the absence of early standardization. 

\begin{table*} [t]
\centering
\caption{Impact of CMA Approaches in Mobile Computing.} \label{Impacts}
\begin{tabular}{|c|c|}

\hline   Advantages & Disadvantages \\ 
\hline   Empowered Processing & Dependency to High Performance Networking Infrastructure \\ 
\hline   Prolonged Battery & Excessive Communication Overhead and Traffic\\ 
\hline   Expanded Storage &  Unauthorized Access to offloaded Data \\ 
\hline   Increased Data Safety &   Application Development Complexity\\ 
\hline   Ubiquitous Data Access  & Paid Infrastructures\\ 
and Content Sharing & \\ 
\hline   Protected Offloaded Contents &  Inconsistent Cloud Policies and Restrictions \\ 
\hline   Enriched User Interface &  Service Negotiation and Control\\ 
\hline   Enhanced Application Generation & Nil \\ 
\hline 
\end{tabular} 
\end{table*}

\subsubsection{Prolonged Battery}
Long-lasting battery can be considered as one the most significant achievements of CMA approaches for large number of mobile users. Smartphone manufacturers have already utilized high speed, multi-core ARM processors (e.g., Cortex-A57 Processor\footnote{http://www.arm.com/products/processors/cortex-a50/cortex-a57-processor.php}) being able to perform daily computing needs of mobile end-users. However, such giant processing entities consume large energy and quickly drain the battery that irks end-users. CMA solutions can noticeably save energy \cite{Altamimi2012} by migrating heavy and energy-intensive computing to the cloud for execution. Although energy efficiency is one of the most important challenges of current CMA systems, several efforts such as \cite{DiFrancesco2012,Heide2012,Smith2012,Liu2011_VM,Gao2012} are endeavoring to comprehend the energy implications of exploiting cloud-based resources from mobile devices and shrinking their energy overhead.

In traditional cyber foraging or surrogate computing approaches, energy is saved by computation offloading, but several issues such as lack of mobility support and resource elasticity can neutralize the benefits of energy-hungry task offloading.

\subsubsection{Expanded Storage} \label{cloudstorage}
Infinite cloud storage accessible from smartphones enables users to utilize large number of applications and digital data on device. Hence, they are not obliged to frequently install and remove popular applications and data due to the space limit. Online connectivity is essential to access cloud storage. In such online storage systems, data are manually or automatically updated to the online storage for maintaining the consistency of the online storage system. Storing applications in cloud storage provides the opportunity to update the code without consuming any mobile I/O transactions which enhances user experience and improves the smartphones' energy efficiency \textemdash because I/O transactions are energy-hungry tasks. 

\subsubsection{Increased Data Safety}
CMA efforts can bring the benefit of data safety to the mobile users. Naturally, stored data on mobile devices are susceptible to loss, robbery, physical damage, and device malfunction. Storing sensitive and personal data such as online banking information, online credentials, and customer related information on such a risky storage significantly degrades the quality of user experience and hinders usability of mobile devices. Due to the scarce computing resources, especially energy in mobile devices, performing complex and secure encryption provisions is not feasible. Hence, by storing data in a reliable cloud storage \cite{Zeng2009, Zheng2010}, users ensure data availability and safety regardless of time, place, and unforeseen mishaps. Threats such as device robbery or physical damage to the mobile devices will effect on the tangible value of the device rather than intangible value of the data.

\subsubsection{Ubiquitous Data Access and Content Sharing}
Cloud infrastructures play a vital role in enhancing data access quality. Storing data in cloud resources enables mobile users to access their digital contents anytime, anywhere, from any device. Hence, the impact of temporal, geographical, and physical differences is noticeably decreased that enriches user experience.

Moreover, cloud storages facilitate data sharing and contribution among authorized users. Every file and folder in cloud, usually has a protected unique access link that can be obtained by the owner to share them among legitimate users. Network traffic is hence, shrunk because data is accumulated in a central server accessible to unlimited users from various machines. Cloud can significantly enhance data transfer among different mobile devices. One of the most irksome user's impediments is to transfer data from current mobile device to a new handset. Apart from its temporal cost, porting data from one device to another, especially to a heterogeneous device is a risky practice that puts data is in the risk of corruption and loss of integrity. Stored data on Cloud remain safe and can be synchronized to any number of mobile devices with minimum risk. However, a reliable data access control mechanism is required to adjust user permissions. 

\subsubsection{Protected Offloaded Content}
Cloud storage solutions aim to protect remote codes and data while ensure user's privacy. This is one of the most important gains of replacing surrogates with cloud resources. Cloud servers deploy virtualization technology to isolate the guest environment from other guests and also from their permanent software stack. Moreover, cloud vendors deploy strict security and privacy policies to not only ensure confidentiality of user content, but also to protect their properties and business. Implement internal security provisions particularly the state-of-the-art biometric security systems to protect their physical infrastructures and avoid unauthorized access. Employing complex content encryption, frequent patching, and continuous virus signature update inside the company premise or seeking technical services from a trusted third party \cite{Yang2011a} are other examples of security provisions undertaken in cloud to further protect cloud storage.  

\subsubsection{Enriched User Interface}
As described in \ref{visualization}, visualization shortcomings of mobile devices diminish user experience and hinder smartphones' usability. However, cloud resources can be exploited to perform intensive 2D or 3D screen rendering. The final screen image can be prepared based on the smartphone screen size and streamed to the device. Consequently, screen adaptation also is achieved when cloud side processing engine automatically alter the presentation technique to match screen image with the device screen size.

\subsubsection{Enhanced Application Generation}
Cloud resources and cloud-based application development frameworks similar to $\tcmu$Cloud and CMH, facilitate application generations in heterogeneous mobile environment. Once a cloud component is built, it can be utilized to develop various distributed mobile applications for large number of dissimilar mobile devices. In the presence of cloud components, application programmer needs to develop native mobile components and integrate them with relevant, prefabricated cloud components to develop a complex application. When a mobile-cloud application is developed for Android device, by slightly changing native components the application is transited to new OS like iOS\footnote{\url{http://www.apple.com/ios/}} and Symbian\footnote{\url{http://licensing.symbian.org/}} which significantly save time and money.

\subsection{Disadvantages}
Despite of many advantageous aspects of cloud services, their success is hindered by several drawbacks and shortcomings that are discussed as follows.

\subsubsection{Dependency to High Performance Networking Infrastructure}
CMA approaches demand converged wired and wireless networking infrastructures and technologies to fulfill intersystem communication requirements. In wireless domain, CMAs need high performance, robust, reliable, high bandwidth wireless communication to realize the vision of computing anywhere, anytime, from any-device. In wired communication, fast reliable communications ground is essential to facilitate live migration of heavy data and computations to a regional cloud-based resources near the mobile users. Efforts such as next generation wireless networks \cite{Nasser2006} and the open mobile infrastructure \cite{OWA2011} with Open Wireless Architecture (OWA) by Sieneon \cite{sieneon} contribute toward enhancing the networking infrastructures' performance in MCC.

\subsubsection{Excessive Communication Overhead and Traffic}
Mobile data traffic is significantly growing by ever-increasing mobile user demands for exploiting cloud-based computational resources. Data storage/retrieval, application offloading, and live VM migration are example of CMA operations that drastically increase traffic leading to excessive congestion and packet loss. Thus, managing such overwhelming traffic and congestion via wireless medium becomes challenging, especially when offloading mobile data are distributed among helping nodes to commute to/from the cloud. Consequently, application functionality and performance decrease leading to user experience degradation.

\subsubsection{Unauthorized Access to Offloaded Data}
Since cloud clients have no control over their remote data, users contents are in risk of being accessed and altered by unauthorized parties. Migrating sensitive codes as well as financial and enterprise data to publicly accessible cloud resources decreases users’ privacy, especially enterprise users. Moreover, storing business data in the cloud is likely increasing the chance of leakage to the competitor firm. Hence, users, especially enterprise users hesitate to leverage cloud services to augment their smartphones.

\subsubsection{Application Development Complexity}
The excessive complexity created by the heterogeneous cloud environment increases environment's dynamism and complicates mobile application development. Mobile application developers are required to acquire extensive knowledge of cloud platforms (i.e., cloud OSs, programming languages, and data structures) to integrate cloud infrastructures to the plethora of mobile devices. Understanding and alleviating such complexity impose temporal and financial costs on application developers and decrease success of CMA-based mobile applications.
\begin{figure*}[t]
\centering
\includegraphics[scale=0.48]{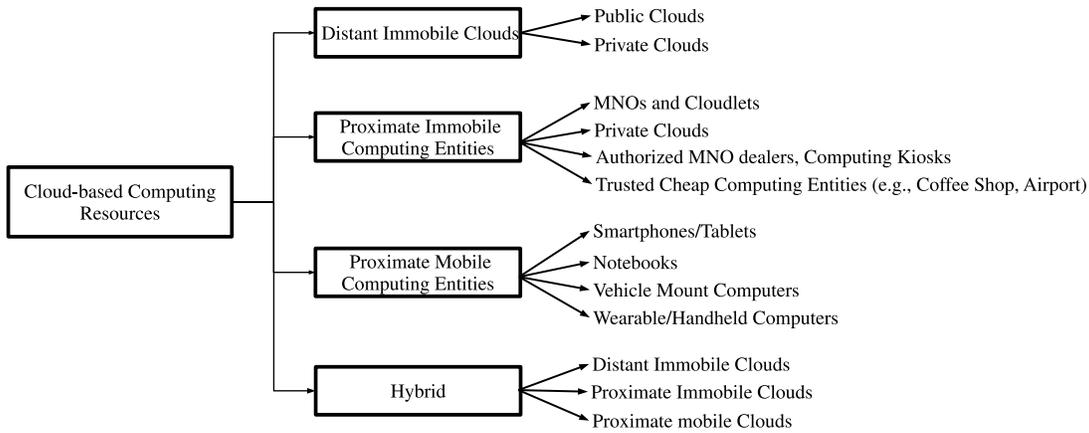}
\caption{Taxonomy of Cloud-based Computing Resources.}
\label{fig:Cloud-basedresources}
\end{figure*}
\subsubsection{Paid Infrastructures}
Unlike the free surrogate resources, utilizing cloud infrastructure levies financial charges to the end-users. Mobile users pay for consumed infrastructures according to the SLA negotiated with cloud vendor. In certain scenarios, users prefer local execution or application termination because of monetary cloud infrastructures cost. However, user payment is an incentive for cloud vendors to maintain their services and deliver reliable, robust, and secure services to the mobile users.

In addition, cloud vendors often charge mobile users twice; once for offloading contents to the cloud and once again when users decide to transfer their cloud data to another cloud vendors to utilize more appropriate service (e.g., monetary and QoS (Quality of Service) aspects). 

\subsubsection{Inconsistent Cloud Policies and Restrictions}
One of the challenges in utilizing cloud resources for augmenting mobile devices is the possibility of changes in policies and restrictions imposed by the cloud vendors. Cloud service providers apply certain policies to restrain service quality to a desired level by applying specific limitations via their intermediate applications like Google App Engine bulk loader\footnote{https://developers.google.com/appengine/docs/python/tools/uploading data}. Services are controlled and balanced while accurate bills will be provided based on utilized resources. 

Also, service provisioning, controlling, balancing, and billing are often matched with the requirements of desktop clients rather than mobile users \cite{Kuhne2012}. Considering the great differences in wired and wireless communications, disregarding mobility and resource limitations of mobile users in design and maintenance of cloud can significantly impact on feasibility of CMA approaches. Hence, it is essential to amend restriction rules and policies to meet MCC users’ requirements and realize intense mobile computing on the go.

\subsubsection{Service Negotiation and Control}
While cloud users are required to negotiate and comply with the cloud terms and conditions for a certain period of time, often cloud agreements are nonnegotiable and policies might change over the time. Moreover, there is no control over the cloud performance and commitments in the absence of a controlling authority or a trusted third party. Hence, CMA services are always volatile to the service quality of cloud vendors.

\section{Taxonomy of Cloud-based Computing Resources}\label{cloud-based resources}
Researchers  \cite{Huerta-Canepa, cuervo2010maui,Satyanarayanan2009, Verbelen2012,Hung2012,Kempa, Kosta2011, Guo2011, A.Manjunatha2010,Zhang2011, Chun2011, Chun2009,March2011,Luo2009, Badidi2011, Liu2009, Kumar2010,Chuna,Lu2011,Kemp2010a, MOMCC, SAMI,Ma2012, Verbelen2012,Gu2012,Giurgiu2009} aim to obtain user requirements and preferences by exploiting varied types of cloud-based resources to augment computing capabilities of resource-constraint smartphones.  Based on the distance and mobility traits of such varied cloud-based computing resources, we classify them into four groups, namely distant immobile clouds, proximate immobile computing entities, proximate mobile computing entities, and hybrid that are taxonomized in Figure \ref{fig:Cloud-basedresources} and explained as follows. Table \ref{cloud-basedservers} represents the comparison results of these cloud-based computing resources. This Table can be utilized as a guideline for appropriate selection of cloud-based infrastructures in future CMA researches.  

\subsection{Distant Immobile Clouds}
Public and private clouds comprised of large number of stationary servers located in vendors or enterprises premises are classified in this category. They are highly available, scalable, and elastic resources that are often located far from the mobile nodes accessible via the Internet. Although public cloud resources are likely more secure compared to the other types of resources due to complex security provisions and on-premise infrastructures \cite{PANDA,Kamara2010,wang2009ensuring, mather2009cloud}, they are vulnerable to security attacks and breaches like Amazon EC2 crash \cite{CHRISTIANCACHIN2011} and Microsoft Azure security glitch \cite{microsoftAzureCrash}. Accessing cloud resources, especially public clouds often carries the risk of communicating through the risky channel of Internet. However, giant clouds are endeavoring to maintain security -for more market share- and could establish high reputation-based trust by providing long-term services to the users.
\begin{table*} [tbph]
\centering
\caption{Comparison of Cloud-Based Servers.} \label{cloud-basedservers}
\begin{tabular}{|c|c|c|c|c|}\hline
& Distant clouds& Proximate immobile & Proximate mobile & Hybrid \\  
& & computing entities & computing entities& \\ \hline
Architecture &\multicolumn{4}{c|}{Distributed}\\ \hline
Ownership& Service provider& Public & Individual& Hybrid\\  \hline
Environment& Vendor Premise & Business Center & Urban Area & Hybrid\\ \hline
Availability & High & Medium& Medium& High \\ \hline
Scalability& High & Medium & Medium& High\\ \hline
Sensing Capabilities& Medium & Low & High & High \\ \hline 
Utilization Cost& \multicolumn{4}{c|}{Pay-As-You-Use}\\ \hline
Computing Heterogeneity & High & Medium& High&High \\ \hline
Computing Flexibility& High&Medium&High& High\\ \hline
Power Efficiency & High&Medium&Medium& High\\ \hline
Execution Performance& High&Medium &  Medium & High\\ \hline
Security and Trust&\multicolumn{1}{c|}{High} & \multicolumn{1}{c|}{Moderate}&\multicolumn{1}{c|}{Low}&\multicolumn{1}{c|}{High}\\ \hline
Utilization Rate & \multicolumn{4}{c|}{High}\\ \hline
Execution Platform & VM & VM & Physical/VM& Physical/VM \\ \hline
Resource Intensity & High & Moderate& Moderate& Rich\\ \hline
Complexity & Low & Moderate& Moderate& High\\ \hline
Communication Technology&3G/WiFi& WiFi & WiFi & 3G/WiFi\\ \hline
Communication Latency & High & Low & Low& Moderate\\ \hline
Execution Latency & Low & Medium & Medium & Low \\ \hline
Maintenance Complexity & Low & Medium& Medium& High \\ \hline
\end{tabular}
\end{table*}

Additionally, the performance and efficacy of these approaches are affected by long WAN latency due to the long distance between mobile client and stationary cloud data centers. One potential approach to shorten the distance between mobile device and cloud is to migrate the remote code and data to the computing resources near to the mobile device via live migration of the VM from the cloud \cite{LiveVM2005}. However, live migration of VM is a non-trivial task that requires great deal of research and development, particularly in networking environment due to several issues such as large VM size, hard-to-predict user mobility pattern, and limited, intermittent wireless bandwidth.

Resource utilization is enhanced in clouds due to the virtualization technology deployment. Several VMs can be executed on a single host to increase the utilization efficiency of the clouds, while each computation task runs on a single isolated VM loaded on a physical machine. However, VM security attacks such as VM hopping and VM escape \cite{Owens} can violate the code and data security. VM hopping is a virtualization threat to exploit a VM as a client and attack other VM(s) on the same host. VM escape is the state of compromising the security of the hypervisor and control all the VMs.

\subsection{Proximate Immobile Computing Entities}
The second type of cloud-based computing resources involves stationary computers located in the public places near the mobile nodes. The number of computers in public places such as shopping malls, cinema halls, airports, and coffee shops is rapidly increasing. These machines are hardly performing tense computational tasks and are mostly playing music, showing advertisement, or performing lightweight applications. Moreover, they are connected to the power socket and wired Internet. Therefore, it is feasible to leverage such abundant resources in vicinity and perform extensive computation on behalf of resource-constraint mobile devices. It can also reduce latency and wireless network traffic while increases resource utilization toward green computing. Another group of proximate immobile computers are Mobile Network Operators (MNO) and their authorized dealers scattered in urban and rural areas, private clouds, and public computing kiosk \cite{Garriss2008} that can be exploited in smartphone augmentation. 

\begin{figure*}
\centering
\includegraphics[scale=0.2]{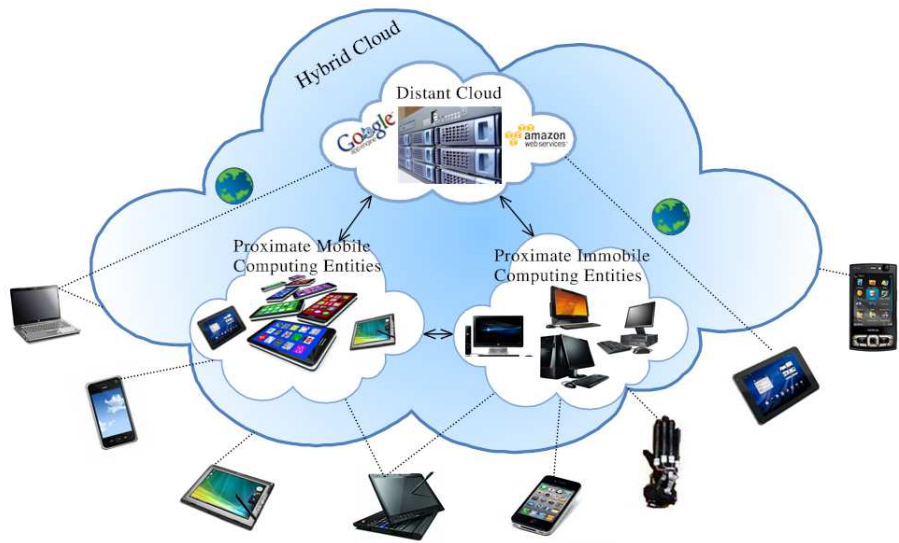} 
\caption{The Hybrid Cloud Concept for MCC.} \label{hybrid}
\end{figure*}

However, protecting security and privacy of mobile user and computer owner hinder utilization of such nearby resources. Several shortcomings such as insufficient on-premise security infrastructure, lake of tight security mechanisms, and inefficient update and maintenance procedures inhibit utilizing such resources (except MNOs) for CMA approaches. Owners of these resources may attack mobile users and access their private data on the mobile devices or falsify offloading results. Also, malicious users may leverage these resources as an attacking point to violate mobile users' security and privacy. On the other hand, security and privacy of resource owners are also susceptible to violation. Owners of computer devices participating in resource sharing require robust mechanisms to protect and isolate the guest code and data from their host applications and data. Virtualization aims to realize such isolation mechanism, but issues such as VM hopping and VM escape require to be addressed before its successful adoption \cite{Owens}. Among all proximate immobile resources, MNOs may be considered unique in terms of security and privacy features. MNOs, in general, have been serving mobile users for long time and could establish high degree of trust among mobile users. It is feasible to assume that MNO's certified dealers also can inherit MNO's trust if central management and monitoring process is undertaken by MNOs \cite{SAMI}.

\subsection{Proximate Mobile Computing Entities}
In this category of cloud-based infrastructures, various mobile devices, particularly smartphones, Tablets, notebooks, wearable computers, and handheld computing devices play the role of servers based on cloud computing principles. The main benefit of utilizing nearby mobile resources is their proximity to the mobile clients. Also, hardware and platform heterogeneity \cite{Z2012} between mobile servers and clients can be mitigated, because both sides are mainly ARM-based devices with mobile OSs. Moreover, contemporary smartphones are able to provide value added context- and social-aware services \cite{Lane2010,Lukowicz2012} that contribute to the context-awareness of mobile applications. However, mobile devices' resources are limited and they are unable to perform intensive context-computing \cite{Makris2013}. Realizing distributed computing on cluster of nearby mobile devices requires several issues, particularly application architectures, resource scheduling, and mobility to be addressed.

Moreover, security and privacy of mobile devices as a service provider is a critical concern in CMA. Mobile devices are intrinsically susceptible to loss and robbery, and their constraint resources inhibit exploiting robust security mechanisms inside the device. Furthermore, with ever-increasing popularity of mobile Apps (i.e., mobile applications) in online App stores such as Google Play\footnote{\url{https://play.google.com/store}} and Samsung APPs\footnote{\url{http://www.samsungapps.com}} \cite{Shen2012} number of mobile security threats are rising sharply and malware-contaminated Apps are becoming serious threats to the mobile users \cite{ANIA2012}. Several security threats have been identified in an experiment of Android mobile applications with the potential to violate the security of mobile users \cite{Enck}. Risk of such contaminated codes can likely be transferred to the non-contaminated mobile devices by utilizing their computation resources and request for results of a remote computation. Hence, establishing trust between mobile devices and end-users becomes a challenging task.

\subsection{Hybrid (Converged Proximate and Distant Computing Entities)}
Hybrid infrastructures as depicted in Figure \ref{hybrid} are comprised of various proximate and distant computing nodes, either mobile or immobile. The main idea behind building hybrid resources is to employ heterogeneous computing resources to create a balance between user requirements (mainly latency and computation power) and available options \cite{Rahimi2012}. The latency sensitive codes are offloaded to the nearest computing device(s) whereas the most intensive and least latency sensitive tasks are migrated to the furthest resources. Perhaps, the utilization costs of nearby resources are more than the remote servers. 

Beneficial characteristics of hybrid resources summarized in Table \ref{cloud-basedservers} advocates their usefulness in maximizing the augmentation benefits. However, deployment, management, and resource scheduling processes in dynamic mobile environment are non-trivial tasks. Developing an autonomic management system similar to CometCloud \cite{Kim2011} in cloud computing and MAPCloud \cite{Rahimi2012} in MCC to automatically manage, optimize, and adapt hybrid infrastructures in the cloud-mobile applications can significantly improve the quality of hybrid CMA approaches.

Hybrid cloud infrastructures can deliver enhanced security and privacy features to the CMA approaches and increase the QoS. Hybrid clouds are comprised of resources with varied security, privacy, and trust features which can be efficiently utilized by CMA and mobile users as a trade-off. For instance, security sensitive computations can perform a security-latency trade-off and execute computation inside a secure distant cloud.

\begin{figure*}[t]
\centering
\includegraphics[scale=0.23]{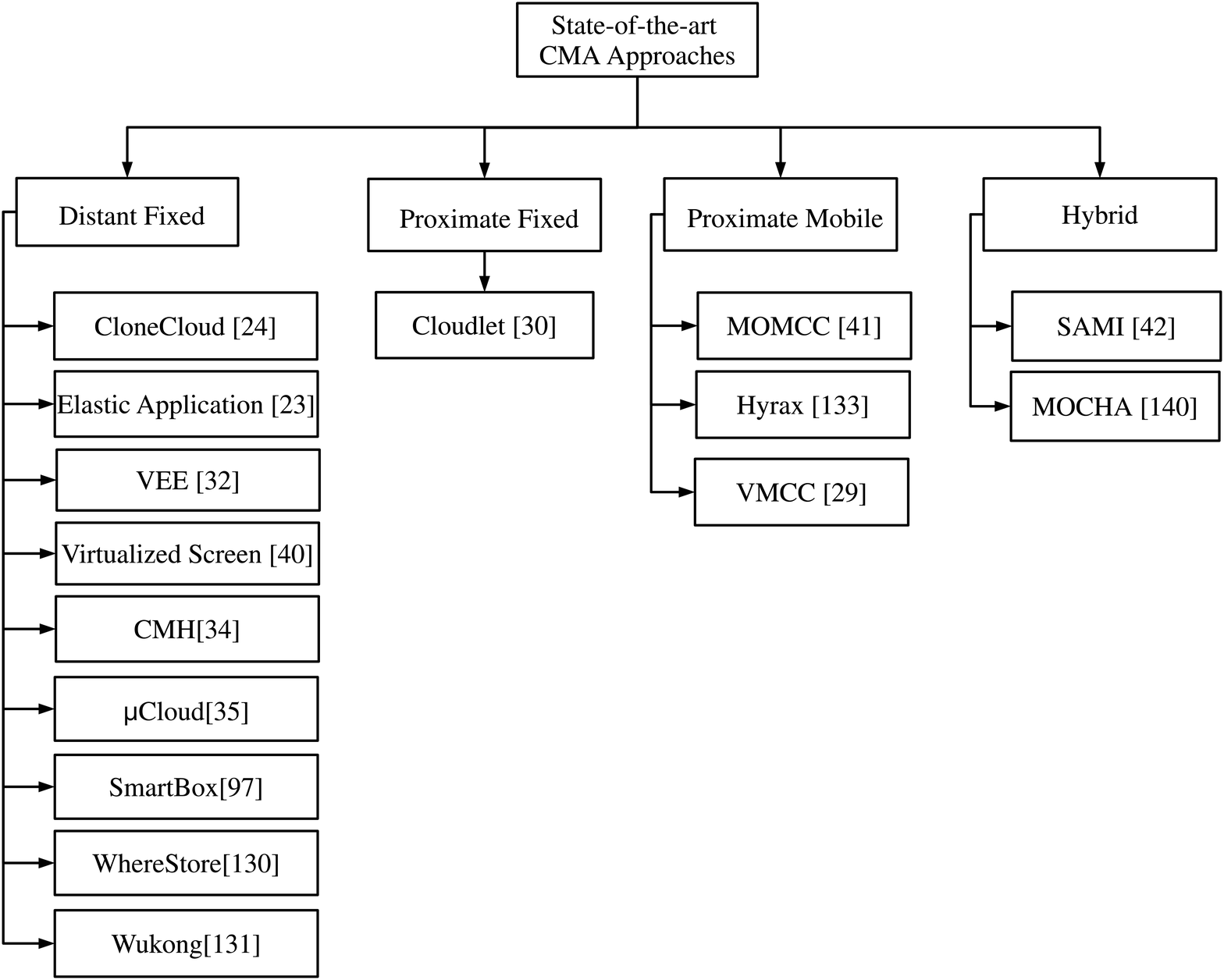} 
\caption{Taxonomy of State-of-the-art CMA Models.} \label{taxonomyCMA}
\end{figure*}

\section{The State-of-the-art CMA Approaches: Taxonomy}\label{state-of-the-art}
\textit{Cloud-based Mobile Augmentation (CMA) is the-state-of-the-art mobile augmentation model that leverages cloud computing technologies and principles to increase, enhance, and optimize computing capabilities of mobile devices by executing resource-intensive mobile application components in the resource-rich cloud-based resources.} According to our resource classifications in previous Section, we analyze and taxonomize the state-of-the-art CMA approaches into four models, namely distant fixed, proximate fixed, proximate mobile, and hybrid which are depicted in Figure \ref{taxonomyCMA}. For each model, we describe few CMA efforts and tabulate the comparison results in Figure \ref{table colored}. 

\subsection{Distant Fixed}
Majority of CMA approaches \cite{Chuna, Chun2009, Kemp2010, Kemp2010a, Kempa,Guo2011, Zhang2011, cuervo2010maui,Chun2011,Verbelen2012, Broberg2009, DiFrancesco2012} leverage fixed cloud infrastructures in distance due to its straightforward approach. Utilizing stationary cloud eliminates several management complexities (e.g., resource discovery and scheduling for mobile cloud-based servers) and alleviates reliability and security concerns \cite{Kristensen2010}. Works in this class of CMA systems aim at reducing the complexity and overhead of utilizing distant cloud. For instance, in \cite{DiFrancesco2012} authors propose an energy-efficient offline job scheduling  model based on makespan minimization model to enhance energy efficiency of distant fixed CMA systems. Their main notion is to separate the data transmission from the job execution. During their work, authors provide several optimization solutions aiming to reduce the energy consumption of the device during the offloading process. However, for the sake of simplicity, the authors study the energy consumption of tasks in offline mode only which does not consider runtime dynamism of MCC.

Exploiting cloud resources is feasible in several real scenarios such as live cloud streaming \cite{Lawton2012}, enterprise applications (e.g., Customer Relation Management (CRM) and enterprise resource planning \cite{Hariharan2008}), and Social Networking. Cloud streaming mechanism has already described in \ref{cloud-streaming} as an example of utilizing distance fixed resources. In \cite{Hariharan2008}, researchers leverage cloud resources in developing a CRM application to enhance efficiency of sale representatives for a pharmaceutical company. The representative meets the physician in medical centers to promote drugs, present samples and promotions material, and he records all sale results and details through the mobile application. The huge database of the company is stored inside the cloud and the sale representative can request to process, get, or update data in database without storing data locally. 

We describe some of the distant fixed CMA approaches that utilize distant fixed cloud resources for mobile augmentation as follows. The terms immobile, fixed, and stationary are interchangeably used with the same notion.

\begin{itemize}
\item \textit{CloneCloud:}
CloneCloud \cite{Chun2011}  is a cloud-based, fine-grained, thread-level, application partitioner and execution runtime that clones entire mobile platform into the cloud VM and runs the mobile application inside the VM without performing any change in the application code. The CloneCloud enables local execution of remaining mobile application when remote server is running the intensive components unless local execution tries to accessing the shared memory state. Cloud resources in this effort simulate distributed execution of a monolithic application in a resourceful environment without engaging application developer into the distributed application programming domain. CloneCloud  can significantly reduce the overall execution time using thread-level migration. When the local execution reaches the intensive component(s), the CloneCloud system offloads the component(s) to the cloud and continues local execution until the application fetches data from the migrated state. The local execution is paused until the results are returned and integrated to the local application. 

However, the communication overhead of transferring the clone of mobile platform, application, and memory state and frequent synchronization of the shared data between the mobile and cloud can shrink the power of cloud. Such overhead becomes more intense in case of heavy, data- and communication-intensive, and tightly coupled mobile applications where an alternative execution of resource-intensive and lightweight components exists. Frequent code and data encapsulation and migration, and mobile-cloud data synchronization excessively increase the communication traffic and impact on execution time and energy efficiency of the offloading. 

\item \textit{Elastic Application:}
Elastic application model \cite{Zhang2011} is a CMA proposal leverages distant fixed cloud data center for executing resource-intensive components of the mobile application. Authors in this model partition a mobile application into several small components, called weblet. Weblets are created with least dependency to each other to increase system robustness while decrease the communication overhead and latency. The weblet execution is dynamically configured to either perform locally or remotely, based on the weblet's resource intensity, execution environment quality, and offloading objectives.

The distinctive attribute of this proposal is that application execution can be distributed among more than one machine and cooperative results can be pushed back to the device. To achieve such goal, multiple elasticity patterns namely replication, splitter, and aggregator are defined. In replication pattern, multiple replicas of a single interface are executed on multiple machines inside the cloud. Hence, failure in one replica will not compromise the system performance. In splitter pattern, the interface and implementation are separated so that several weblets with varied implementations can share a single interface. In aggregator, the results of multiple weblets are aggregated and pushed to the device for optimized accuracy and efficacy. 

The authors endeavor to specify the execution configurations (specifying where to run the weblets) at runtime to match the requirements of the applications and users. To enhance the overall execution performance and enrich user experience, the system is able to run the weblets both locally and remotely. A weblet can be executed remotely in a low-end device while the same can be executed locally on a high-end device. 

Elastic application model pays more attention to the user preferences by enabling different running modes of a single application (e.g., high speed, low cost, offline mode). However, it engages application developers to determine weblets organization based on the functionality, resource requirements, and data dependency. But, the characteristics of the weblets are mainly inherited from the well-known web services to decrease the programmer burdens. 

\item \textit{Virtual Execution Environment(VEE):}
Hung et al. \cite{Hung2012} propose a cloud-based execution framework to offload and execute the intensive Android mobile applications inside the distant cloud's virtual execution environment. The quality and accuracy of execution environment is highly influenced by the comprehensiveness and accuracy of emulated platform. This method uses a software agent in both mobile and cloud sides to facilitate the overall system management. The agent in mobile device initiates VM creation and clones the entire application (even native codes and UI components) and partial data/memory state from device to the cloud. Unlike CloneCloud, VEE aims to reduce latency by migrating the segment of data stack explicitly created and owned by the application to the VM instead of copying the entire memory; cloning the entire memory state, especially for heavy applications significantly increases latency and traffic. 

During remote execution, the system frequently synchronizes the changes between device and cloud to keep both copies updated. In order to increase the quality and efficiency of remote execution in virtual environment and avoid data input loss at application suspension stages, the system stores input events (reading a file, capturing a face, storing a voice) exploiting a record/replay scheme and pseudo checkpoint methods. However, these methods engage application developers to separate the application state into two states, namely global and local and to specify the global data structures. The global state contains the program domain and application flow, whereas the local state contains local data structures required by a method. Programmer usually needs to identify global state when the application is paused. Once the application is suspended, the global state will be loaded to avoid re-execution and the latest Android checkpoint is applied to the system to reflect all the changes made from the last checkpoint. However, all changes, especially user input might be lost from the last checkpoint. To record the changes after the last checkpoint, the record/replay mechanism is deployed by creating a pseudo checkpoint. To create a pseudo checkpoint, the application notifies the local agent to identify the input events and record required information. Upon the application resumption, the pseudo checkpoint is restored to restore the application to the state prior to the suspension.

In this effort, code security inside the cloud is enhanced by exploiting encryption and isolation approaches that protects offloaded code from cloud vendors eavesdropping. Using probabilistic communication QoS technique,  this is aimed to provide a communication-QoS trade-off. For instance, the control data (usually small volume) needs highest accuracy while video streaming data (often large volume) requires less communication accuracy. Moreover, the authors are optimistic that offering secondary tasks such as automatic virus scanning, data backup, and file sharing in the virtual environment can enhance quality of user experience.

Although this approach aims to enhance the quality of application execution and augment computation capability of mobile clients and save energy, but responsiveness in interactive applications are likely low due to remote UI execution. Instead of migrating entire application to the cloud, it might be more beneficial to utilize some of the local mobile resources instead of treating mobile device as a dump client. Data passing between mobile device and cloud for interactive applications might degrade quality of experience, especially in low-bandwidth, intermittent networks.

\item \textit{Virtualized Screen:}
Virtualized screen \cite{Lu2011} is another example of CMA approaches that aims to move the screen rendering process to the cloud and deliver the rendered screen as an image to the mobile device. The authors aim to enrich the user experience and migrate the screen rendering tasks to the cloud with the assumption that majority of computation- and data-intensive processing take place in the cloud. Hence, abundant cloud resources' exploitation simplifies the CMA system architecture, prolongs mobile battery, and enhances the interaction and responsiveness of mobile applications toward rich user experience. Screen virtualization technique (running partial rendering in cloud and rest in mobile depending on the execution context) is envisioned to optimize user experience, especially for lightweight, high-fidelity, interactive mobile applications that entirely run on local resources. Their conceptual proposal aims to enhance visualization capability of mobile clients, mitigate the impact of hardware and platform heterogeneity, and facilitate porting mobile applications to various devices (e.g., smartphone, laptop, and IP TV) with different screens. 

To reduce the mobile-cloud data transmission, a frame-based representation system is exploited to forward the screen updates from the cloud to the mobile. Frame-based representation system captures and feeds the whole screen image to the transmission unit. This approach updates each frame based on the previous frame stored inside both the mobile and cloud. However, a rich interactive, responsive GUI needs live streaming of screen images which is impacted by communication latency. Although the authors describe optimized screen transmission approaches to reduce the traffic, the impact of computation and communication latency is not yet clear, as this is a preliminary proposal. Moreover, utilizing virtualized screen method for developing lightweight mobile-cloud application is a non-trivial task in the absence of its programming API.

\item \textit{Cloud-Mobile Hybrid (CMH) Application:}
Unlike application offloading solutions, authors in this proposal \cite{A.Manjunatha2010} introduce a new approach of utilizing cloud resources for mobile users. In this effort, the authors propose a novel CMH application model, in which heavy components are developed for cloud-side execution, whereas lightweight or native codes are developed for mobile devices execution. CMH Applications execution does not need profiling, partitioning, and offloading processes and hence produce least computation overhead on mobile devices. Upon successful cloud-side execution, the results are returned back to the mobile for integrating to the native mobile components. 

However, developing CMH applications is significantly complex due to the interoperability and vendor lock-in problems in clouds and fragmentation issue in mobiles \cite{Z2012}. Cloud components designed for a specific cloud are not able to move to another cloud due to underlying heterogeneity among clouds. Similarly, mobile components developed for a particular platform cannot be ported to different platforms because of heterogeneity. Yet isolating development of mobile and cloud components creates further versioning and integration challenges.

To mitigate the complexity of CMH application developments and facilitate portability, the authors leverage Domain Specific Language (DSL) \cite{Ranabahu2011,Deursen2000}. A DSL is a programming language with major focus on solving problem in specific domains. MATLAB\footnote{http://www.mathworks.com/products/matlab/} is a well-known DSL-based tool for mathematicians. A parser takes a DSL script and converts codes into an in-memory object to be forwarded to various automatic component generators. The system needs different code generators for various mobile and cloud platforms. Once the mobile and cloud components are generated, the CMH application can be assembled for various mobile-cloud platforms. However, utilizing DSL-based techniques requires more generalization efforts to be beneficial in developing  all types of CMH applications.

\item \textit{$\tcmu$Cloud:}
Similar to the CMH framework, $\tcmu$Cloud \cite{March2011} is a modular, mobile-cloud application framework that aims to facilitate mobile-cloud application generation, promote application portability, minimize the development complexity, and enhance offline usability in intensive mobile-cloud applications. Fulfilling separation of concerns vision, skilled programmers independently develop self-contained components which do not have any direct inter communications with each other. Unskilled mobile users can mash-up (assembling available components to build complex application) these prefabricated components to generate a complex mobile-cloud application. Cloud vendors provide infrastructure and platform as cloud services to run prefabricated cloud components. The main idea in this proposal is to avoid local execution of the resource-intensive components. Hence, components are identified as cloud, mobile, and hybrid; mobile components are executable exclusively on mobile and cloud components are strictly developed for cloud server while hybrid components can either run locally or remotely. Hybrid components have either multiple implementations or a single implementation that need a middleware for execution. Each component has a triplet of identifier, input/output parameters, and configuration.

To alleviate offline usability issue, the authors leverage mobile-side queuing and cloud-side caching to maintain data in case of disconnection. Data will be transferred upon reconnection. Application is partitioned into components and organized as a directed graph. Nodes represent components and vertices indicate data/control flow.  Application is divided into three fragments; in each fragment, a managing unit called orchestrator executes and maintains component's mash-up process. The output of each component is forwarded using the pass-by-value semantic as an input to the subsequent component. 

Unlike Elastic Application model, the design and implementation of components in $\tcmu$Cloud is statically performed in early development phase. Thus, any improvement in resource availability of mobile devices or environmental enhancement (like bandwidth growth) will not improve the overall execution of $\tcmu$Cloud applications. Such inflexibility decreases the application execution performance and degrades the quality of user experience.

\item \textit{SmartBox:}
Smartbox \cite{Zheng2010} is a self-management, online, cloud-based storage and access management system developed for mobile devices to expand device storage and facilitate data access, and sharing. It is a write-once, read many times system designed to store personal data such as text, song, video, and movies which is not appropriate for large scale computational datasets. In Smartbox, mobile devices are associated with a shadow storage to store/retrieve personal data using a unique account. To facilitate data sharing among larger group of end-users in office or at home, a public storage space is provisioned.

Smartbox exploits traditional hierarchical namespace for smooth navigation and employs an attribute-based method to facilitate data navigation and service query using semantic metadata such as the publisher-provider metadata. Data navigation and query using tiny keyboard and small screen irk mobile users when inquiring and navigating stored data in cloud. However, mobile users need always-on connectivity to access online cloud data which is not yet achieved and is unlikely to become reality in near future. 

\item \textit{WhereStore:}
WhereStore \cite{Stuedi2010} is a location-based data store for cloud-interacting mobile devices to replicate necessary cloud-stored mobile data on the phone. The main notion in this effort is that users in different places doing various activities need dissimilar types of information. For instance, a foreign tourist in Manhattan requires information about nearby places of interest rather than all the country. Hence, identifying the location and caching predicted data deemed can enhance the system efficiency and user experience. However, efficient prediction of future user location and required data, and determining the right time for caching data are challenging tasks.

\item \textit{Wukong:}
Wukong \cite{Mao2011} is a cloud oriented file service for multiple mobile devices as a user-friendly and highly available file service. The authors provision support of heterogeneous back-end services such as FTP, Mail, and Google Docs Service in a transparent manner leveraging a service abstraction layer (SAL). Wukong enables applications to access cloud data without being downloaded into the local storage of mobile device.

Authors introduce cache management and pre-fetch mechanisms in different scenarios to increase performance while decreasing latency. However, it cannot always reduce latency due to the bandwidth limitation and I/O overhead. In operations with long gap between open and read, it is beneficial to pre-fetch data from cloud to the device that significantly improves user experience. Data security is enhanced via an encryption module and bandwidth is saved using a compression module. While compression methods utilized in this proposal is inefficient for multimedia files like image and music, it can compress text and log files noticeably. 

We conclude that one of the most effective solutions to tackle bandwidth and latency limitations in CMA approaches, especially cloud storage is to decrease the volume of data using imminent compression methods. While various compression methods work well on specific file types, a cognitive or adaptive compression method with focus on multimedia files can significantly improve the feasibility of cloud-storage systems.
\end{itemize}

\subsection{Proximate Fixed}
Researchers have recently proposed CMA approaches in which nearby stationary computers are utilized. Utilizing nearby desktop computers initiates new generation of services to the end-user via mobile device. In \cite{Satyanarayanan2009}, the authors provide a real scenario in which Ron, a patient diagnosed with Alzheimer, receives cognitive assistance using an augmented-reality enabled wearable computer. The system consists of a lightweight wearable computer and a head-up display such as Google Glass\footnote{\url{http://www.google.com/glass/start/}} equipped with a camera to capture the environment and an earphone to send the feedback to the patient. The system captures the scene and sends the image to the nearby fix computers to interpret the scene in the image using the object or face recognition, voice synthesizer, and context-awareness algorithms. When Ron looks at a person for few seconds, the person's name and some clue information is whispered in Ron's ear to help greeting with the person. When he looks at his thirsty plant or hungry dog, the system reminds Ron to irrigate the plant and feed his dog. The nearby resources are core component of this system to provide low-latency real-time processing to the patient. In this part, we explain one of the most prominent proximate fixed efforts as follows.

\begin{figure*} [t]
\centering
\includegraphics[scale=0.34]{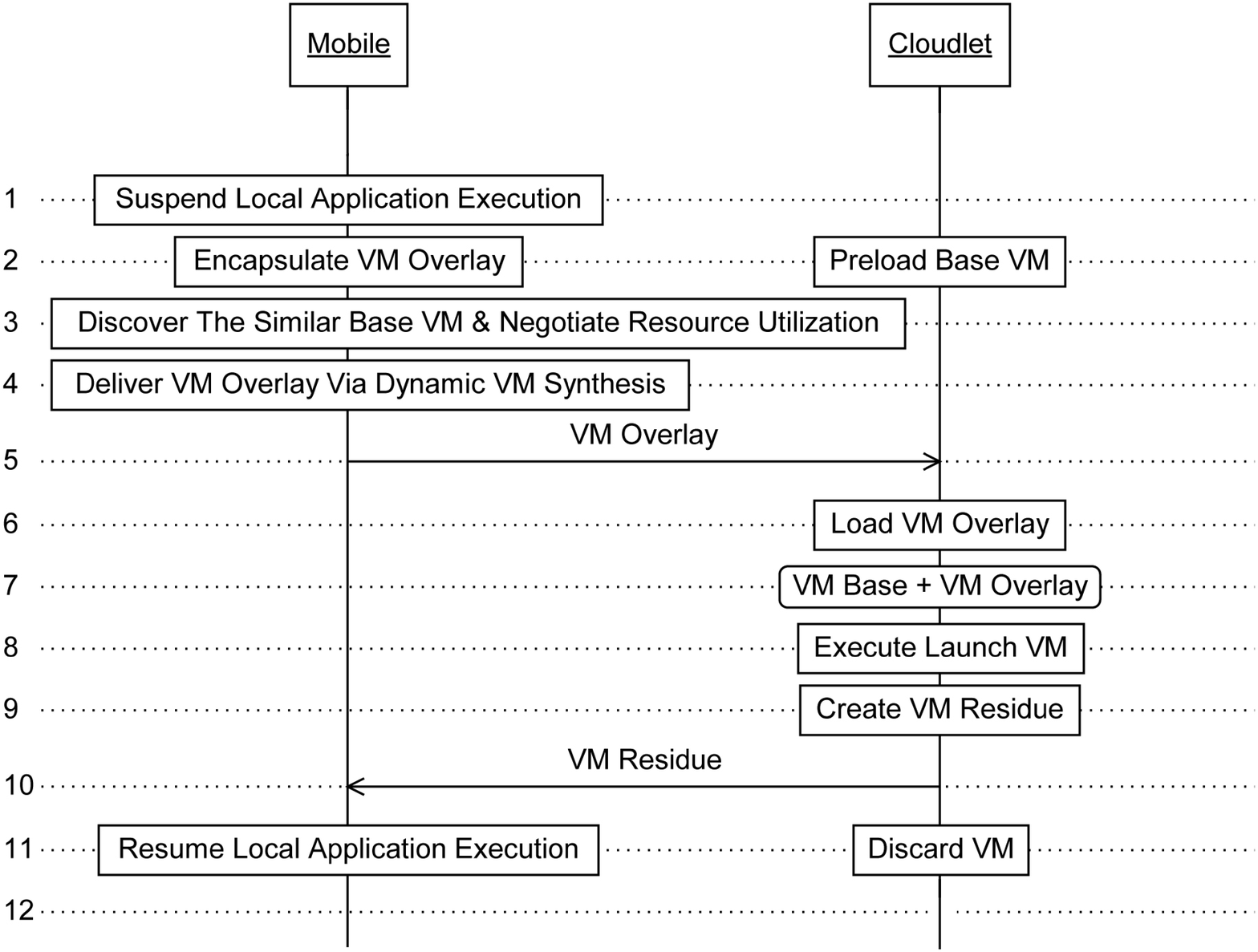}
\caption{Cloudlet-based Resource-Rich Mobile Computing Life Cycle.} \label{sequencediagramcloudlet}
\end{figure*}

\begin{itemize}
\item \textit{Cloudlet:}
Cloudlet \cite{Satyanarayanan2009} is a proximate immobile cloud consists of one or several resource-rich, multi-core, Gigabit Ethernet connected computer aiming to augment neighboring mobile devices while minimizing security risks, offloading distance (one-hop migration from mobile to Cloudlet), and communication latency. Mobile device plays the role of a thin client while the intensive computation is entirely migrated via Wi-Fi to the nearby Cloudlet. Although Cloudlet utilizes proximate resources, the distant fixed cloud infrastructures are also accessible in case of Cloudlet scarcity. The authors employ a decentralized, self-managed, widely-spread infrastructure built on hardware VM technology. Cloudlet is a VM-based offloading system that can significantly shrink the impact of hardware and OS heterogeneity between mobile and Cloudlet infrastructures. 

To reduce the Cloudlet management and maintenance costs while increasing security and privacy of both Cloudlet host and mobile guest, a method called ``transient Cloudlet customization'' is deployed which uses hardware VM technology. It enables Cloudlet customization prior to the offloading and performs Cloudlet restoration as a post-offloading cleanup process to restore the host to its original software stake. The VM encapsulates the entire offloaded mobile environment (data state and code) and separates it from the host permanent software. Hence, feasibility of deploying Cloudlet in public places such as coffee shops, airport lounge, and shopping malls increases. 

Unlike CloneCloud and Virtual Execution Environment efforts that migrate the entire mobile OS clone to the cloud, Cloudlet assumes that the entire OS clone exists and is preloaded in the host and runs on an isolated VM. In mobile side, instead of creating the VM of the entire mobile application and its memory stack, the systems encapsulates a lightweight software interface of the intensive components called VM overlay. 

The VM overall offloading performance is further enhanced by exploiting Dynamic VM Synthesis (DVMS) method since its performance solely depends on the mobile-Cloudlet bandwidth and cloudlet resources. DVMS assumes that the base VM is already available in the target Cloudlet and user can find the matching execution environment (VM base) among silo of nearby Cloudlets. Upon discovery and negotiation of the Cloudlet, the DVMS offloads the VM overlay to the infrastructure to execute launch VM (base + overlay). Henceforth, the offloaded code starts execution in the state it was paused. Upon completion of Cloudlet execution the VM residue is created and sent back to the mobile device. In the Cloudlet, the VM is discarded as a post-offloading cleanup process to restore the original Cloudlet state. In mobile side, the results will be integrated to the application and local execution will be resumed. To present a clear understanding of the overall process, the sequence diagram of Cloudlet-based resource-rich mobile computing is depicted in Figure \ref{sequencediagramcloudlet}.

Despite the noticeable offloading improvements in the Cloudlet, its success highly depends on the existence of plethora of powerful Cloudlets containing popular mobile platforms' base VM. Encouraging individual owners to deploy such Cloudlets in the absence of monetary incentives is an issue that must be addressed before deployment in real scenarios. Although energy efficiency, security and privacy, and maintenance of Cloudlet are widely acceptable, further efforts are required to protect the overall CMA process. Moreover, few minutes offloading latency in Cloudlet is unacceptable to users \cite{Tolia2006}.

\end{itemize}

\subsection{Proximate Mobile}
Recently, several researchers \cite{hyrax, ChongleiMei2011, MOMCC, Gao2012, Mei2012, Huerta-Canepa,Guirguis2011} propose CMA approaches in which nearby mobile devices lend available resources to other mobile clients for execution of resource-intensive tasks in distributed manner. Utilizing such resources can enhance user experience in several real scenarios such as Optical Character Recognition (OCR) and natural language processing applications. The feasibility of utilizing nearby mobile devices is studied in \cite{Huerta-Canepa} where Peter, a foreign tourist, visiting a Korean exhibition and finds interest in an exhibit, but cannot understand the Korean description. He can take a photo of the manuscript and translate it using the OCR application, but his device lacks enough computation resources. Although he can exploit the Internet web services to translate the text, the roaming cost is not affordable to him. Hence, he leverages a CMA solution by utilizing computation resources of nearby mobile devices to complete the task. Some of the CMA efforts whose remote resources are proximate mobile devices are explained as follows.
\begin{figure*} [t]
\centering
\includegraphics[scale=0.22]{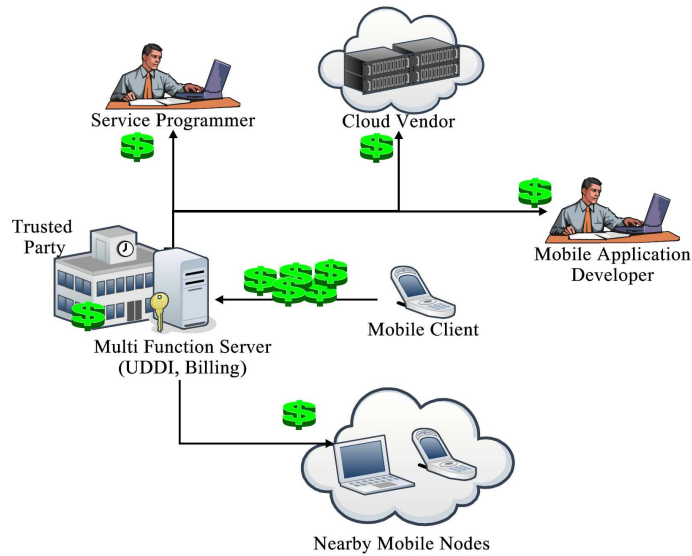}
\caption{MOMCC Concept.} \label{MOMCC}
\end{figure*}
\begin{itemize}
\item \textit{MOMCC:}
Market-Oriented Mobile Cloud Computing (MOMCC) \cite{MOMCC} is a mobile-cloud application framework based on Service Oriented Architecture (SOA) that harnesses a cluster of nearby mobile devices to run resource-intensive tasks. In MOMCC, mobile-cloud applications are developed using prefabricated building blocks called services developed by expert programmers. Service developers can independently develop various computation services and uploaded them to a publicly accessible UDDI (Universal Description Discovery and Integration) such as mobile network operators.

Services are mostly executed on large number of smartphones in vicinity which can share their computation resources and earn some money. To enhance resource availability and elasticity, distant stationary cloud resources are also available if nearby resources are insufficient. In order to become an IaaS (Infrastructure as a Service) provider, mobile devices register with the UDDI and negotiate to host certain services after secure authentication and authorization. Mobile users at runtime contact UDDI to find appropriate secure host in vicinity to execute desired service on payment. The collected revenue is shared between service programmer, application developer, UDDI, and service host for promotion and encouragement. Figure \ref{MOMCC} depicts the MOMCC concept.

However, MOMCC is a preliminary study and its overall performance is not yet evident. Several issues are required to be addressed prior to its successful deployment in real scenarios. Executing services on mobile devices is a challenging task considering resource limitation, security, and mobility. Also an efficient business plan that can satisfy all engaging parties in MOMCC is lacking and demand future efforts. MOMCC can provide an income source for mobile owners who spend couple of hundred dollars to buy a high-end device. In addition, faulty resource-rich mobile devices that are able to function accurately can be utilized in MOMCC instead of being e-waste.

\item \textit{Hyrax:}
Hyrax \cite{hyrax} is another CMA approach that exploits the resources from a cluster of immobile smartphones in vicinity to perform intense computations. Hyrax alleviates the frequent disconnections of mobile servers using fault tolerance mechanism of Hadoop. Similar to MOMCC and Cloudlet, due to resource limitations of smartphone servers, the accessibility to distant stationary clouds is also provisioned in case the nearby smartphone resources are not sufficient. However, Hyrax does not consider mobility of mobile clients and mobile servers. Hence, deployment of Hyrax in real scenarios may become less realistic due to immobilization of mobile nodes. Lack of incentive for mobile servers also hinders Hyrax success. 

Hyrax is a MCC platform developed based on Hadoop \cite{White2012} for Android smartphones. In developing Hyrax, the MapReduce \cite{Dean2004} principles are applied utilizing Hadoop as an open source implementation of MapReduce. MapReduce is a scalable, fault-tolerant programming model developed to process huge dataset over a cluster of resources. Centralized server in Hyrax runs two client side processes of MapReduce, namely NameNode and JobTracker processes to coordinate the overall computation process on a cluster of smartphones. In smartphone side, two Hadoop processes, namely DataNode and TaskTracker are implemented as Android services to receive computation tasks from the JobTracker. Smartphones are able to communicate with the server and other smartphones via 802.11g technology. 

Nevertheless, the cloud storage connectivity in Hyrax is missing. It demands several gigabytes of local storage to store data and computation. Hence, user cannot access distributed data over the Internet or Ethernet. The author utilizes the constant historical multimedia data to avoid file sharing. Hence, it is less beneficial for interactive and event-oriented applications whose data frequently changes over the execution and also data-intensive applications that require huge database. The overall overhead in Hyrax is high due to the intensity of Hadoop algorithm which runs locally on smartphones. 

\item \textit{Virtual Mobile Cloud Computing (VMCC):}
Researchers in \cite{Huerta-Canepa} aim to augment computing capabilities of stable mobile devices by leveraging an ad-hoc cluster of nearby smartphones to perform intensive computing with minimum latency and network traffic while decreasing the impact of hardware and platform heterogeneity. During the first execution, required components (proxy creation and RPC support) are added to the application code to be used for offloading; the modified code will remain for future offloading. For every application,  the system determines the number of required mobile servers, security and privacy requirements, and offloading overhead. The system continuously traces the number of total mobile servers and their geolocation to establish a peer-to-peer communication among them.  Upon decision making the application is partitioned into small codes and transferred to the nearby mobile nodes for execution. The results will be reintegrated back upon completion.

However, several issues encumber VMCC's success. Firstly, this solution, similar to Hyrax, is not suitable for a moving smartphone since the authors explicitly disregard mobility trait of mobile clients. Secondly, every computing job is sent to exactly one mobile node; so, the offloading time and overhead will be increased when the serving node leaves the cluster. Thirdly, the offloading initiation might take long since the offloading's overall performance highly depends on the number of available nearby nodes; insufficient number of mobile nodes defers offloading. Finally, in the absence of monetary incentive for mobile nodes the likelihood of resource sharing among resource-constraint mobile devices is low. 
\end{itemize}

\subsection{Hybrid}
Hybrid CMA efforts are budding \cite{SAMI,Soyata2012,Rahimi2012} to optimize the overall augmentation performance and researchers are endeavoring to seamlessly integrate various types of resources to deliver a smooth computing experience to mobile end-users. For instance, mCloud \cite{Bahl2012} is an imminent proposal to integrate proximate immobile and distant stationary computing resources. Authors are aiming to enable mobile-users to perform resource-intensive computation using hybrid resources (integrated cloudlet-cloud infrastructures). Hybrid solutions aim to provide higher QoS and richer interaction experience to the mobile end-users of real scenarios explained in previous parts. For instance, in the foreign tourist example, the image can be sent to the nearby mobile device of a non-native local resident for processing. When the processing fails due to lack of enough resources, the picture can be forwarded to the cloud without Peter pays high cost of international roaming (Peter may pay local charge). 

We review some of the available hybrid CMAs as follows. 

\begin{itemize}
\item \textit{SAMI:} 
SAMI (Service-based Arbitrated Multi-tier Infrastructure for mobile cloud computing) \cite{SAMI} proposes a multi-tier IaaS to execute resource-intensive computations and store heavy data on behalf of resource-constraint smartphones. The hybrid cloud-based infrastructures of SAMI are combination of distant immobile clouds, nearby Mobile Network Operators (MNOs), and cluster of very close MNOs’ authorized dealers depicted in Figure \ref{SAMI}. The compound three level infrastructures aim to increase the outsourcing flexibility, augmentation performance, and energy efficiency. The MNO's revenue is hiked in this proposal and energy dissipation is prevented. Nearby dealers can be reached by Wi-Fi. MNO's can be accessed either directly via cellular connection or through dealers via Wi-Fi and broadband. Connection is established via cellular network to contact distant stationary clouds.  The cluster of nearby stationary machines (MNO dealers located in vicinity) performs latency-sensitive services and omits the impact of network heterogeneity. SAMI leverages Wi-Fi technology to conserve mobile energy because it consumes less energy compared to the cellular networks \cite{Perrucci}. In case of nearby resource scarcity or end-user mobility, the service can be executed inside the MNO via cellular network. However, if the resources in MNO are insufficient, the computation can be performed inside the distant immobile cloud. 

The resource allocation to the services is undertaken by arbitrator entity based on several metrics, particularly resource requirements, latency, and security requirements of varied services. The arbitrator frequently checks and updates the service allocation decision to ensure high performance and avoids mismatch.

\begin{figure} [t]
\centering
\includegraphics[scale=0.24]{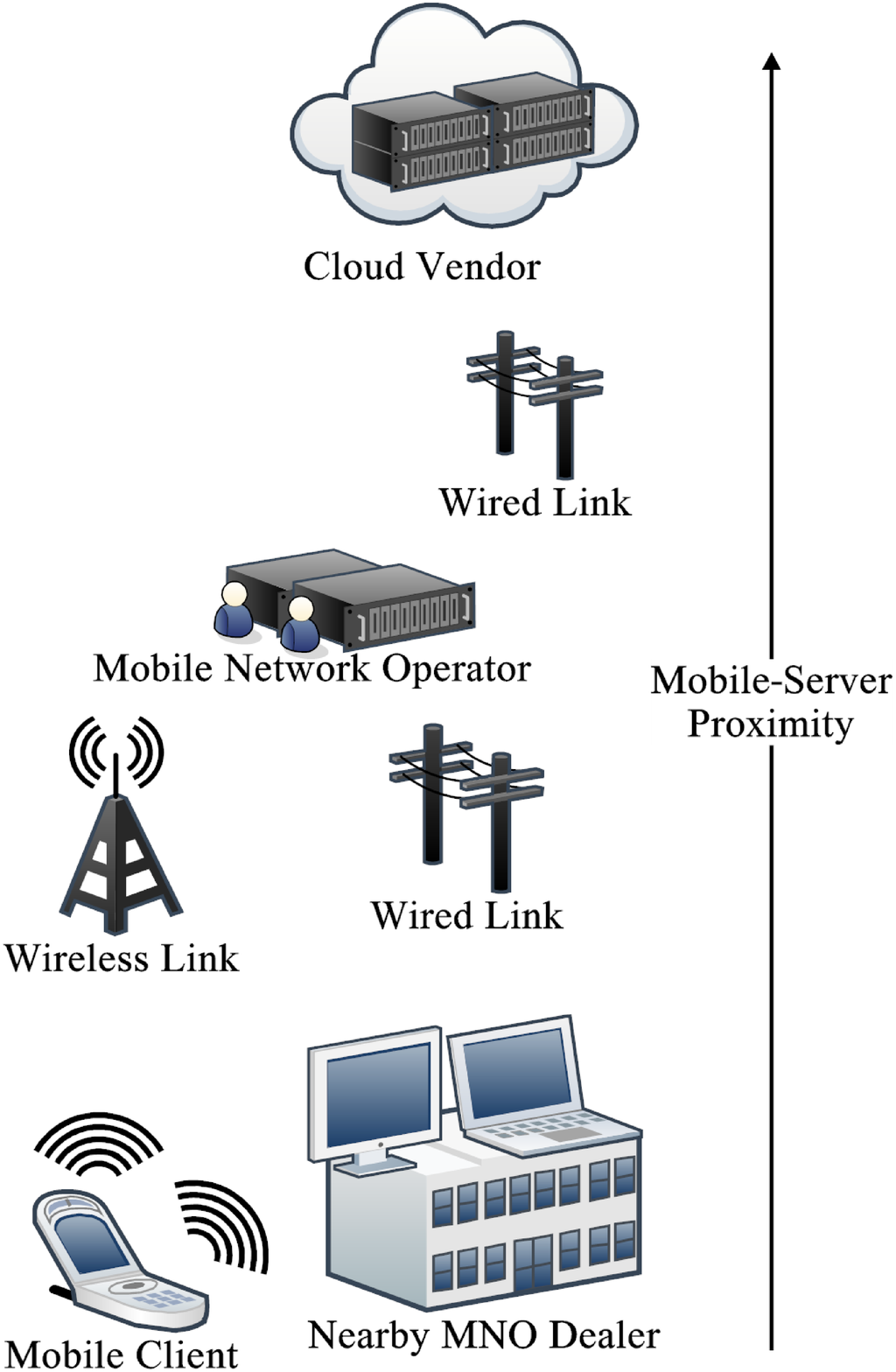}
\caption{SAMI: A Multi-Tier Cloud-based Infrastructure Model.} \label{SAMI}
\end{figure}

To enhance security of infrastructures, SAMI employs comparatively reliable and trustworthy entities, namely clouds, MNOs, and MNO trusted dealers. MNOs have already established reputation-trust among mobile users and can enforce a strict security provisions to establish indirect trust between dealers and end users ensuring that user's security and privacy will not be violated. SAMI application development framework facilitates deployment of service-based platform-neutral mobile applications and eases data interoperability in MCC due to utilization of SOA.

However, SAMI is a conceptual framework and deployment results are expected to advocate its feasibility. SAMI  imposes a processing overhead on MNOs due to continuous arbitration process. Deployment, management, and maintenance costs of SAMI are also high due to the existence of various infrastructure layers. Moreover, though the authors discuss the monetary aspects of the proposal, a detailed discussion of the business plan is missing, for example in what scenario resource outsourcing is affordable for the mobile application? How does income should be divided among different entities to be satisfactory? 
\begin{figure*} [tbph]
\centering
\includegraphics[scale=0.88]{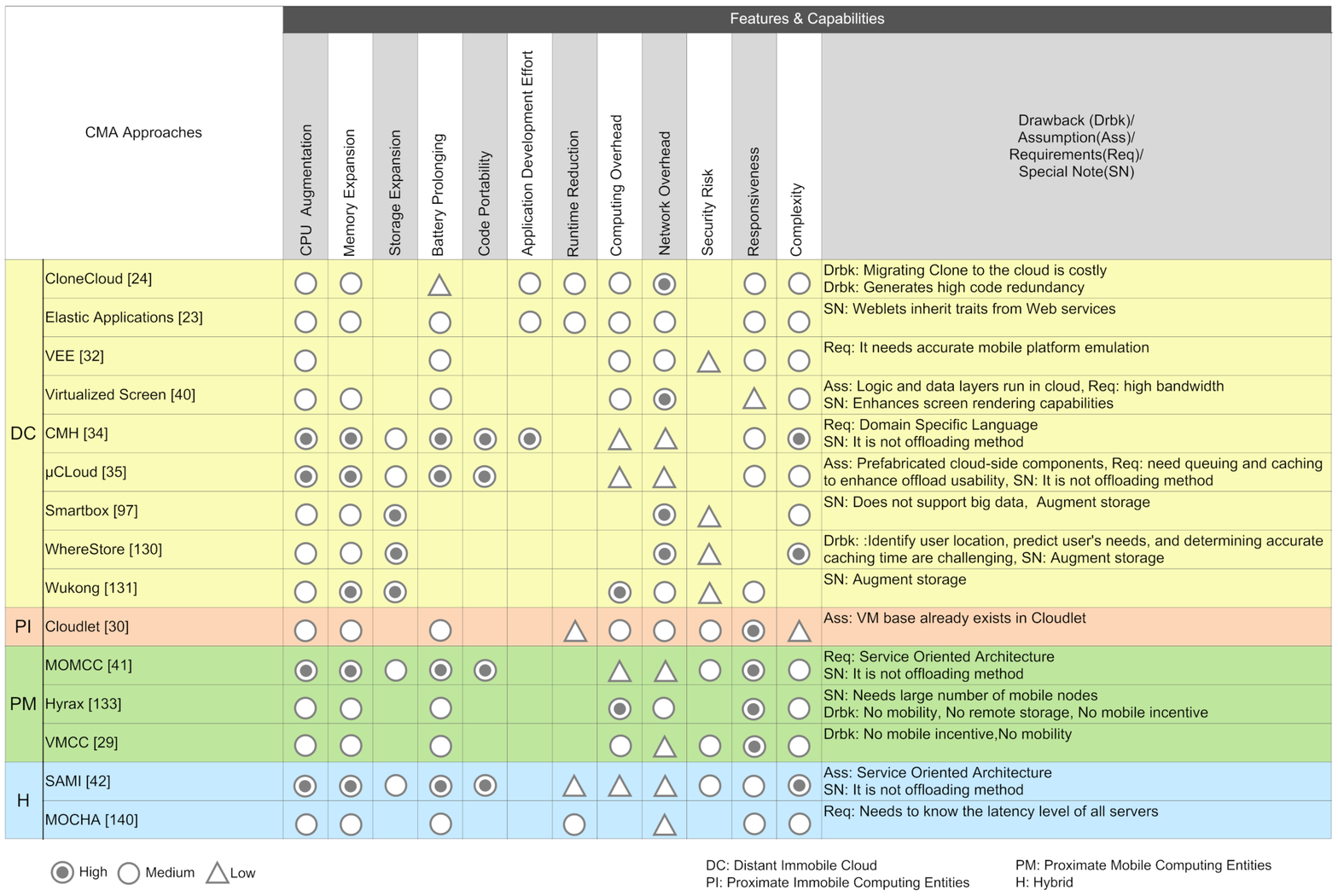}
\caption{Comparison of CMA Approaches.} \label{table colored}
\end{figure*}

\item \textit{MOCHA:} In MOCHA \cite{Soyata2012} authors propose a mobile-cloudlet-cloud architecture for face recognition application using mobile camera and hybrid infrastructures of nearby Cloudlet and distant immobile cloud. Cloudlet is a specific, cheap cluster of computing entities like GPU (Graphics Processing Unit) capable of massively processing data and transactions in parallel. Cloudlets are able to be accessed via heterogeneous communication technologies such as Wi-Fi, Bluetooth, and cellular. The mobile often access processing resources via Cloudlet rather than directly connecting to the cloud, unless accessing cloud resources bears lower latency.

Cloudlet receives the smartphones’ intensive computation tasks and partitions them for distribution between itself and distant immobile clouds to enhance QoS  \cite{Satyanarayanan2009}. MOCHA leverages two partitioning algorithms: fixed and greedy. In the fixed algorithm, the task is equally partitioned and distributed among all available computing devices (including Cloudlet and cloud servers), whereas in greedy algorithm, the task is partitioned and distributed among computing devices based on their response times; the first partition is sent to the quickest device while the last partition is sent to the slowest device. The response time of the task partitioned using greedy approach is significantly better than fixed, especially when Cloudlet server is utilized in augmentation process and large number of clouds with heterogeneous response time exist.

However, smartphones in MOCHA require prior knowledge of the communication and computation latency of all available computing entities (Cloudlet and all available distant fixed clouds) which is a resource-hungry and time-consuming task.
	
\end{itemize}

\section{CMA Prospectives}\label{sec:cma-prospectives}
People dependency to mobile devices is rapidly increasing \cite{Survey2011, Gartner2012} and smartphones have been using in several crucial areas, particularly healthcare (tele-surgery), emergency and disaster recovery (remote monitoring and sensing), and crowd management to benefit mankind \cite{Viswanathan2012,Koukoumidis2012,Kranz2012}. However, intrinsic mobile resources and current augmentation approaches are not matching with the current computing needs of mobile-users, and hence, inhibit smartphone's adoption. Upon slow progress of hardware augmentation, the highly feasible solution to fulfill people computing needs is to leverage CMA concept. This Section aims to present set of guidelines for efficiency, adaptability, and performance of forthcoming CMA solutions. We identify and explain the vital decision making factors that significantly enhance quality and adaptability of future CMA solutions and describe five major performance limitation factors. We illustrate an exemplary decision making flowchart of next generation CMA approaches.

\subsection{CMA Decision Making Factors}
\begin{figure*}[t]
\begin{center}
\includegraphics[scale=0.66]{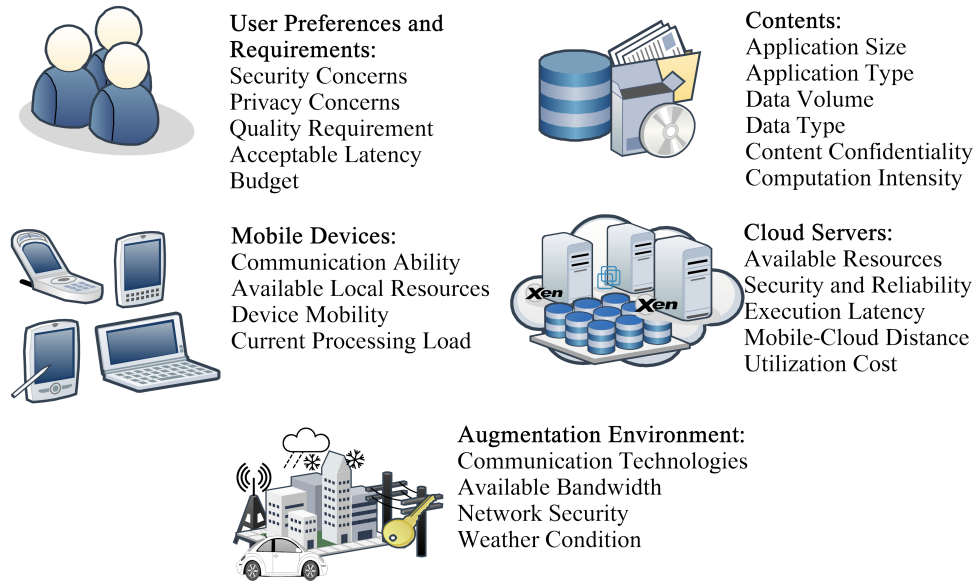}
\caption{Critical Factors in CMA Decision Making.} \label{factor}
\end{center}
\end{figure*}

These factors can be used to decide whether to perform CMA or not and are needed at design and implementation phases of next generation CMA approaches. We categorize the factors into five main groups of mobile devices, contents, augmentation environment, user preferences and requirements, and cloud servers, which are depicted in Figure \ref{factor} and explained as follows.

\subsubsection{Mobile Devices}
From the client perspective, amount of native resources including CPU, memory, and storage is the most important factor to perform augmentation. Also, energy is considered a critical resource in the absence of long-spanning batteries. The trade-off between energy consumed by augmentation and energy squandered by communication is a vital proportion in CMA approaches \cite{Miettinen2010}. Device mobility and communication ability (supporting varied technologies such as 2G,3G,Wi-Fi) are other metrics that are important in the offloading performance. 

\subsubsection{Contents}
Another influential factor for CMA decision making is the contents' nature. The code granularity and size as well as data type and volume are example attributes of contents that highly impact on the overall augmentation process. Hence, the augmentation should be performed considering the nature and complexity of application and data. For instance, latency sensitive small data are efficient to be processed locally, whereas sensitive big data are encouraged to be stored in a large reliable cloud storage. Similarly, offloading a coarse-grained, large code to a distant fixed cloud via a low bandwidth network is not feasible. 

\subsubsection{Augmentation Environment}
Mobile computing is a heterogeneous environment comprised of non-uniform mobile nodes, communication technologies, and  resources. One of the most influential environment-dependent factors is the wireless communication medium in which majority of communications take place. Wireless is an intermittent, unreliable, risky, and blipping medium with significant impact on the quality of augmentation solutions. The overall performance of a low cost, highly available, and scalable CMA approach is magnificently shrunk by the low quality of communication medium and technologies. Selecting the most suitable technology considering the factors like required bandwidth, congestion, utilization costs, and latency \cite{Bellavista2011} is a challenge that affects quality of augmentation approaches in wireless domains. Wireless medium characteristics impose restrictions when specifying remote servers at design time and runtime. 

Moreover, dynamism and rapidly changing attributes of the runtime environment noticeably impact on augmentation process and increase decision making complexity. Augmentation approaches should be agile in dynamic mobile environment and instantaneously reflect to any change. For example, user movement from high bandwidth to a low bandwidth network, receding from the network access point, and rapidly changing available computing resources complicate CMA process.

\subsubsection{User Preferences and Requirements}
End-users' physical and mental situations, individual and corporate preferences, and ultimate computing goals are important factors that affect offloading performance. Some users are not interested to utilize the risky channel of Internet, while others may demand accessing cloud services through the Internet. Hence, users should be able to modify technical and non-technical specifications of the CMA system and customize it according to their needs. For example, user should be able to alter degree of acceptable latency against energy efficiency of an application execution. Selecting the most appropriate resource among available options can also enhance overall user experience. 

\subsubsection{Cloud Servers} \label{servers}
As explained, CMA approaches can leverage various types of cloud resources to enhance computing capabilities of mobile devices. Therefore, the overall performance and credibility of the augmentation approaches highly depend on the cloud-based resources' characteristics. Performance, availability, elasticity, vulnerability to security attacks, reliability (delivering accurate services based on agreed terms and conditions), cost, and distance are major characteristics of the cloud service providers used for augmenting mobile devices. 

Utilizing clouds to augment mobile devices notably reduces the device ownership cost by borrowing computing resources based on pay-as-you-use principle. Such elastic, cost-effective, reliable, and relatively trustworthy resources are embraced by the scholars, industrial organizations, and end-users towards flourishing CMA approaches. 

\subsection{Performance Limitation Factors}
Performance of varied CMA solutions is impacted by several factors. We describe fix major performance limitation factors as follows.
\subsubsection{Heterogeneity}
MCC is a highly heterogeneous environment comprised of three diversified domains of mobile computing, cloud computing, and networking. Although heterogeneity can provide flexibility to the mobile users by providing selection alternatives, it breeds several limitations and challenges, especially for developing multi-tier CMA-based applications \cite{Z2012}. Dissimilar mobile platforms such as Android, iOS, Symbian, and RIM beside diverse hardware characteristics of mobile device inhibit data and application portability among varied mobile devices. Portability is the ability to migrate code and data from one device to another with no/less modification and change \cite{ANSDIT}. Existing heterogeneity in cloud computing including hardware, platform, cloud service policy, and service heterogeneity originates challenges such as portability and interoperability and fragment the MCC domain. 

Network heterogeneity in MCC is the composition of various wireless technologies such as Wi-Fi, 3G, and WiMAX. Mobility among varied network environments intensifies communication deficiencies and stems complex issues like signal handover \cite{Nasser2006}. Inappropriate decision making during the handover process like (i) less appropriate selection of network technology among available candidates and (ii) transferring the communication link at the wrong time, increases WAN latency and jitter that degrade quality of mobile cloud services. Consequently delay-sensitive content and services are degraded \cite{Yan2010} and adoption of CMA approaches are hindered. 

\subsubsection{Data Volume}
Ever-increasing volume of digital contents \cite{Gantz2008} significantly impacts on the performance of CMA approaches in MCC. Current wireless infrastructures and technologies fail to efficiently fulfill the networking requirements of CMA approaches. Storing such a huge data in a single warehouse is often impossible and demands data partitioning and distributed storage that not only mitigates data integrity and consistency, but also makes data management a pivotal need in MCC \cite{Sakr2011}. Applying a single access control mechanism for relevant data in various storage environments is another challenging task that impacts on the performance and adoption of CMA solutions in MCC.
 
\subsubsection{Round-Trip Latency}
Communication and computation latency is one of the most important performance metrics of mobile augmentation approaches, especially when exploiting distant cloud resources. In cellular communications, distance from the base station (near or far) and variations in bandwidth and speed of various wireless technologies affect the performance of augmentation process for mobile devices. Moreover, leveraging wireless Internet networks to offload content to the distant cloud resources creates a bottleneck. Latency adversely impacts on the energy efficiency \cite{Miettinen2010} and interactive response \cite{Lagar-Cavilla} of CMA-based mobile applications due to excessive consumption of mobile resources and raising transmission delays.

Recently, researches \cite{Bifulco2012, Johansson2012} are emerging toward decreasing the networking overhead and facilitating mobility (both node and code mobility) in cloud-based offloading approaches. For example, Follow-Me Cloud \cite{Bifulco2012} aims at enabling mobility of network end-points across different IP subnets. The authors employ the concept of identifier and locator separation of edge networks using OpenFlow-enabled switches. Leveraging the Follow-Me Cloud, mobile nodes can move among access networks without being notified of any change or session disruption. All corresponding nodes that have been communicating with the mobile node can continue their communication without interruption. When the node migrates, its old IP turn to identifier and its new IP address becomes locator address so that all other nodes can keep communication with the moving node. However, for each packet traveling to/from the mobile node, there is an overhead of manipulating  the locator/identifier values. Future improvement and optimization efforts will enhance the CMA systems' performance.
 
 In cloud side, computation latency significantly impacts on the application responsiveness. Researchers study the impact of cloud computation performance on the execution time and vindicate 12X reduction in performance time violation \cite{Guevara2013}. Thus, the increased latency degrades the quality of user experience and adversely impacts on the user-perceived performance of CMA solutions.

\subsubsection{Context Management and Processing} 
Performance of CMA approaches is noticeably degraded by lack of sufficient, accurate knowledge about the runtime environment. Contemporary mobile devices are capable of gathering extensive context and social information such as available remote resources, network bandwidth, weather conditions, and users' voice and gestures from their surrounding environment \cite{Lane2010,Lukowicz2012}. But, storing, managing, and processing large volume of context information (considering MCC environment's dynamism and mobile devices' mobility) on resource-constraint smartphones are non-trivial tasks.

\subsubsection{Service Execution and Delivery}  
SLA as a formal contract between service consumer and provider enforces resource-level QoS (e.g., memory capacity, compute unit, and storage) against a fee, which is not sufficient for mobile users in highly dynamic wireless MCC environment. User-perceived performance in MCC is highly affected by the quality of cloud computations, wireless communications, and local execution. Hence, varied service providers, including cloud vendors, wireless network providers, and mobile hardware and OS vendors need to collaborate and ensure acceptable level of QoS. For successful CMA approaches, comprehensive real-time monitoring process is expected to ensure that engaging service vendors are delivering required services in acceptable level based on the accepted SLA.

\subsection{CMA Feasibility} \label{feasibility}
Although CMA is beneficial and can saves resources \cite{Kumar2010}, several questions need to be addressed before CMA can be implemented in real scenarios. For instance: is CMA always feasible and beneficial? Can CMA save local resources and enhance user experience? What kind of cloud-based resources should be opted to achieve the superior performance? 

\begin{figure}[t]
\begin{center}
\includegraphics[scale=0.15]{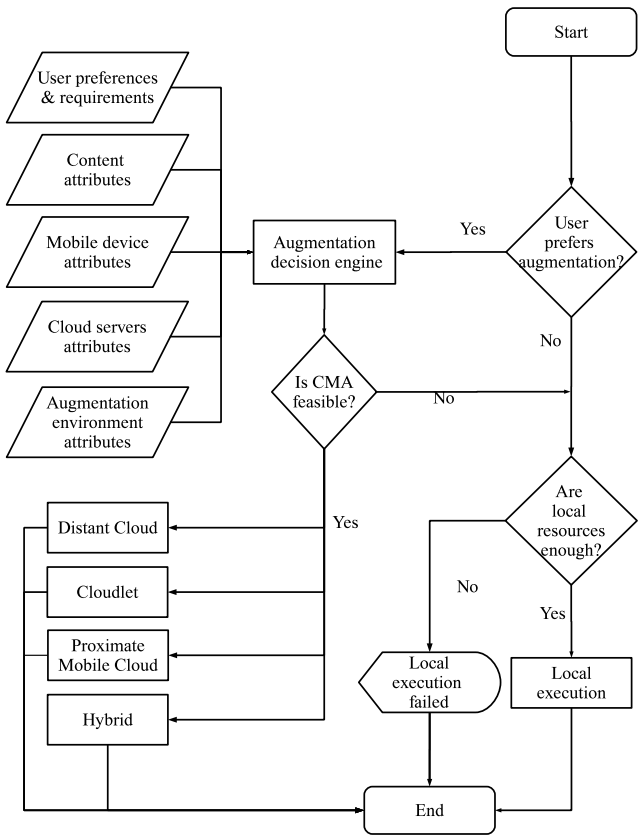}
\caption{An Exemplary Decision Making Flow of Future CMA Approaches.} \label{qualification}
\end{center}
\end{figure}

Vision of future CMA proposals will be realized by accurate sensing and acquiring precise knowledge of decision making factors like user preferences and requirements, augmentation environment, and mobile devices, which are explained in previous part. A decision making system, similar to those used in \cite{cuervo2010maui,Zhang2011,Giurgiu2009}, analyzes these vital factors to determine the augmentation feasibility and specifies if augmentation can fulfill mobile computation requirements and enrich quality of user experience. Figure \ref{qualification} illustrates a possible decision making flow of future CMA approaches.

Availability of mobile resources to manage augmentation process and volume of cloud resources to provision required resources significantly impact on the quality of augmentation \cite{Balan2006}. Similarly, user preferences, limitations, and requirements affect the augmentation decision making. For instance, if augmentation is not permitted by users, the application execution and data storage should be performed locally without being offloaded to a remote server(s) or be terminated in the absence of enough local resources. Similarly, augmentation process can be terminated if the execution latency of delay-sensitive content is sharply increased, quality of execution is noticeably decreased, or security and privacy of users is violated \cite{Sharifi2011}. 

Furthermore, usefulness of CMA approaches highly depends on the execution environment. Offloading computation and mobile-cloud communication ratio, distance from mobile to the cloud, network technologies and coverage, available bandwidth, traffic congestion, deployment cost, and even nature of augmentation tasks alter usefulness of the CMA approaches \cite{Kumar2010}. For instance, performing an offloading method on a data-intensive application (e.g., applying a graphical filter on large number of high quality images) in a low-bandwidth network imposes large latency and significantly degrades user experience which should be avoided. Similarly, migrating a resource-hungry code to an expensive remote resource can be unaffordable practice. Suppose in a sample augmentation approach $R_{C}$ is the total native resources consumed during augmentation, $R_{M}$ is the total native resources consumed for maintenance, and $R_{S}$ is the total resources conserved in augmentation process. Explicitly for a feasible augmentation approach $R_{C} + R_{M} << R_{S}$. However, in some traditional augmentation approaches, the left side of the equation exceeds the right which is not effective in augmenting resource-poor mobile devices \cite{Sharifi2011}. 

\section{Open Challenges} \label{openissues}
In this Section, we highlight some of the most important challenges in deploying and utilizing CMA approaches as the future research directions.

\subsection{Reference Architecture for CMA Development}
Recently, researchers leverage dissimilar structures and techniques in utilizing cloud resources to augment computation capabilities of mobile devices. Also, different efforts focus on varied types of mobile applications, particularly multimedia, intense games, image processing, and workflow processing applications. 

Such diffusion scatters CMA development approaches and increase adaptability challenges of CMA solutions. In the absence of a reference architecture and unified CMA solution, various CMA approaches need to be integrated to all mobile OSs to serve multi-dimensional needs of various mobile users which is a not-trivial task. The reference architecture is expected to be generic enough to be deployed in family of CMA approaches.

\subsection{Autonomic CMA}
CMA approaches are drastically increasing volume of distributed mobile content in a horizontally heterogeneous mobile cloud computing environment \cite{ZohrehSanaei2012}. Exploiting heterogeneous communication technologies to employ diverse cloud-based infrastructures for augmenting a plethora of dissimilar mobile devices is significantly intensifying complexity and management. Employing lateral solutions and controlling the mobile phones using outside entities (e.g., third party management systems) to deal with such complexity might further amplify complexity. A feasible alternative to mitigate such complexity is to develop autonomic self-managing, -healing, -optimizing, and -protecting CMA approaches \cite{Buyya2012} able to adapt to environment dynamism and hide complexity.

\subsection{Application Mobility Provisioning}
Enabling continuous and consistent mobility in CMA models (especially proximate mobile and hybrid) to provision ubiquitous, convenient, on-demand network access to cloud-based computing resources is a vital challenge. Seamless code mobility in CMA models is more challenging compared to the traditional augmentation approaches, because in CMA approaches service providers and consumers can move during the augmentation process which intensifies the code mobility \cite{Yu2012}. Therefore, communication disruption and intermittency can cause several challenges, especially dismissal of always-on connectivity, excessive consumption of limited mobile resources, and frequent interruption of application execution which decreases quality of computing services and degrades quality of mobile user experience \cite{Satyanarayanan2005}. Also, it levies redundant costs on cloud-mobile users and inhibits reliability of CMA models. Hence, alleviating such difficulties using Web advancements\cite{Johansson2012} and imminent lightweight cognitive mobility management systems with least signal traffic and latency can significantly enhance the ubiquitous connectivity and increase the positive impact of CMA.

\subsection{Computing and Temporal Cost of Mobile Distributed Execution}
Noticeable computation and communication cost of migrating tasks from the mobile device to the remote servers and receiving the results is another challenges of CMA approaches in MCC, which is intensified by mobility and wireless communication constraints. Although researchers \cite{MOMCC, hyrax} endeavor to reduce the distance of mobile devices and service providers by leveraging nearby mobile/fixed computing devices, several mechanisms, particularly resource discovery and allocation, service consumer and provider mobility management, and distributed runtime are required to realize the CMA’s vision. Accurate allocation of resources to the mobile computation tasks demands comprehensive knowledge about structure and performance features of available service providers and resource requirements of mobile computation tasks. Thus, QoS-aware scheduling efforts such as \cite{Rahimi2012} are necessary to enhance the CMA usability.
 
\subsection{Seamless Communication} 
Maintaining a continuous communication between mobile service consumers and mobile/fixed service providers in intermittent heterogeneous wireless environment is a non-trivial task. User mobility and wireless disconnection highly impact on resource utilization ratio. When the mobile service providers and consumers loss communication link due to mobility-made prolonged distance, the service consumer requires to either performing local execution or re-initiating augmentation process. Similarly, the resource-constraint mobile server consumes its scarce resources for processing an orphan computation whose results are ineffectual after losing the communication link. Potential solutions may transfer partially-completed tasks to a nearby node or initiate parallel execution on third device before disconnection, or cache results for future references.

\subsection{Multipoint Data Bridging}
Unlike traditional offloading methods that require point-to-point code and data migration and processing, CMA approaches require multipoint data migration and interoperation to achieve the maximum benefits from the distributed heterogeneous infrastructures in mobile-cloud ecosystem. Connecting heterogeneous systems (based on wired or wireless), understanding geographical information resources, and exchanging data between/across two or more heterogeneous systems \cite{Blair2011} are the main issues in CMA system which demand arousing data interoperation techniques in multi-domains mobile cloud environment. The inward heterogeneous architectures and data structures of mobile devices and cloud systems with different APIs can exemplify the intensity of multipoint data bridging challenge \cite{HoganJuly2011}. Hence, offloading computational tasks from a mobile device to a cloud, performing computational interoperation among varied clouds (for cost and performance concerns), and pushing results to the mobile device become challenging tasks \cite{A.Ranabahu}. Therefore, multipoint data bridging in dynamic heterogeneous environment remains as a future research direction to realize accessing, interpreting, processing, sharing, and synchronizing distributed contents.

\subsection{Distributed Content Management}
Rapid growth in digital contents and increasing dependency of mobile users to cloud infrastructures impede content management for mobile users. Researchers distribute code and data among heterogeneous nearby and distant resources via different communication technologies to optimize CMA process. Although executing complex, heavy applications and accessing large data volume are facilitated, managing huge volume of distributed contents for smartphone users is not straightforward. Therefore, enabling mobile users to efficiently locate, access, update, and synchronize highly distributed contents requires future research and developments.

\subsection{Seamless/Lightweight CMA}
Developing lightweight mobile computing augmentation approaches to increase quality of mobile user experience and to develop CMA system independent of any particular situation is a significant challenge in mobile cloud environment. Offloading bulk data in limited wireless bandwidth, and VM initiation, migration, and management in a secure and confidential manner are particular tasks in CMA system that noticeably increase overall execution time, intensify the augmentation latency, and decrease the quality of mobile user experience. CMA approaches are generally hosted and executed inside the mobile devices to conserve their local resources and hence, need to avoid excessive resource hungry transactions.

A feasible approach to decrease the volume of digital contents \textemdash in limited bandwidth networks \textemdash is to utilize effective, efficient data compression methods. Available compression techniques are unlikely efficient considering structure of current multimedia files. Moreover, approaches like Paravirtualiztion \cite{Youseff2006} as a lightweight virtualization technique can reduce the overhead by partially emulating the OS and hardware. Paravirtualization approaches virtualize only parts of the hardware required for computing. Thus, the mobile-side VM creation overhead is diminished and the impact of VM migration on network is reduced. Therefore, realizing lightweight CMA approaches demands lightweight computation and communication techniques (particularly in virtualization, data compression, and encryption methods) to reduce intra-system correspondence, data volume, and I/O tasks.

\subsection{Security in Mobile Cloud}
One of the most challenging aspects of CMA is protecting offloaded code and data inside the cloud. While securing contents inside the mobile consumes huge resources, offloading plain contents through insecure wireless medium and storing plain data inside the cloud highly violates user security and privacy. Despite of the large number of research and development in establishing trust in cloud  \cite{Wang2009f,Wang2009g,mowbray2009client,Ruj2011}, security and privacy is still one of the major user concerns in utilizing cloud resources that impede CMA deployment. Addressing such crucial needs by employing a novel lightweight security algorithm in mobile side and a set of robust security mechanisms in cloud demand future efforts to promote CMA among smartphone users.
In privacy aspects, though recent social behaviors of users in social websites such as Facebook and tweeter advocates that large community of users partially forfeit privacy, they still need certain degree of robust privacy to protect their confidential, clandestine data.

\subsection{Live Virtual Machine Migration}
Live migration of VMs between distributed cloud-based servers (especially for distant servers) is a crucial requirement in successfully adopting CMA solutions in MCC, considering wireless network bandwidth and intermittency, and mobility limitations. When a mobile user moves to a place far from the offloaded contents (code or data), the enlarged distance increases access latency and degrades user-observed application performance. Hence, mobilizing the running VM along with the mobile service consumer without perceivable service interruption becomes essential to avoid user experience degradation. However, sharp growing computation and data volume in blipping wireless environment intensifies live migration of VM. Therefore, efforts similar to VMware vMotion \cite{VMWARECOMPARISION} and \cite{Liu2011_VM, Takahashi2012} are necessary to optimize VM migration in MCC. Reducing computation complexity and overhead, energy, data volume, and communication cost are critical in low-latency low-cost migration of VM in MCC.

Furthermore, after successful live VM migration, it is essential to ensure that migrated VM is seamlessly accessible via initial IP address when the VM changes its physical machine. Future efforts similar to \cite{Watanabe2010, Harney2007} and LISP \cite{Raad2013} are needed to realize the vision of seamless access to migrating VM in MCC.

\section{Conclusions} \label{conclusions}
Augmenting computing capabilities of mobile devices, especially smartphones using cloud infrastructures and principles is an emerging research area. The ultimate goal of CMA solutions is to realize the vision of unrestricted functionality, storage, and mobility regardless of underlying devices and technologies' constraints. Elasticity, availability, and security are considered as the most significant characteristics of remote computing resources. Augmenting smartphones significantly improves their usability and adoption in various critical areas such as healthcare, emergency handling, disaster recovery, and crowd management. Enterprise organizations also leverage such value-added services to enhance quality of services delivering to their end-users. However, several communication and networking challenges, particularly live VM migration, seamless back-end and front-end mobility, handover, context awareness, and location management decelerate adoption of cloud-based augmentation solutions in MCC. Similarly, irreconcilable and unpredictable bandwidth, jitter, communication delay, inconsistent network throughput, and non-elastic wireless spectrum and infrastructures encumber upfront planning for CMA.

In this paper, we presented a comprehensive survey on mobile augmentation process and reviewed several CMA approaches. We described deficiencies of current mobile devices as a motivation for augmentation. Comparing surrogates as traditional servers with contemporary cloud-based infrastructures advocates adaptability of cloud computing and principles as an appropriate back-end technology in mobile augmentation domain. Cloud resources in this domain are manifested as distant immobile clouds, proximate mobile and immobile computing entities, and hybrid combination of resources. CMA approaches not only enhance processing, energy, and storage capabilities of mobile devices, but also amend data safety and security, data ubiquity, accessibility, and user interface while facilitate distributed application programming. Despite of significant CMA approaches advantages, several disadvantages hinder their adoption in mobile computing. Several open challenges particularly, autonomic CMA, seamless application mobility, distributed data management, and multipoint data bridging are the most prominent open challenges that demand future efforts.

\bibliographystyle{IEEEtran}
\bibliography{E:/Dropbox/mandeley/library}

\begin{biography}[{\includegraphics[scale=2.34]{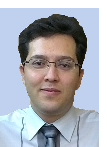}}]{Saeid Abolfazli} is currently a Ph.D. candidate, research assistant in High Impact Research Project (Mobile Cloud Computing: Device and Connectivity) fully funded by Malaysian Ministry of Higher Education, and part time lecturer in the Department of Computer Systems and Technology at the University of Malaya, Malaysia. He received his M.Sc in Information Systems in 2008 from India and BE (Software Engineering) in 2001 from Iran. He has been serving as CEO of Espanta Information Complex during 1999-2006 in Iran. He also was part time lecturer to the ministry of education and Khorasan Technical and Vocational Organization between 2000 and 2006. He is a member of IEEE society and IEEE CS Cloud Computing STC. He has been serving as a reviewer for several international conference and ISI journals of computer science. His main research interests include Mobile Cloud Computing, lightweight protocols, and service oriented computing (SOC). Please write to him at abolfazli.s@gmail.com or abolfazli@ieee.org. For further information, please visit his cyberhome at www.mobilecloudfamily.com
\end{biography}

\begin{biography}[{\includegraphics[scale=0.49]{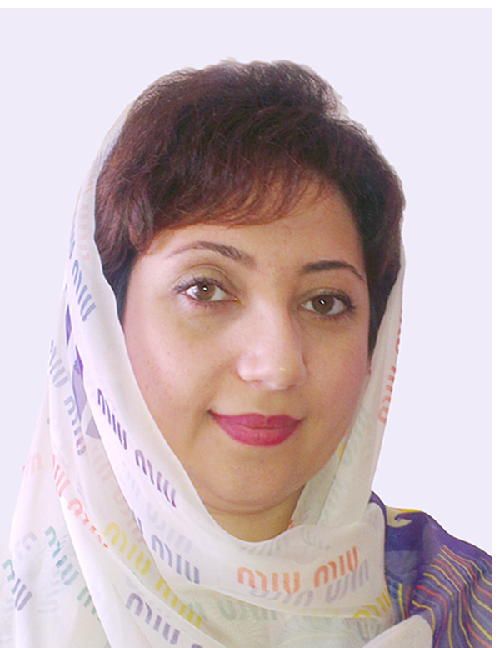}}]{Zohreh Sanaei} is currently a Ph.D. candidate and research assistant in High Impact Research Project (Mobile Cloud Computing: Device and Connectivity) fully funded by Malaysian Ministry of Higher Education in the Department of Computer Systems and Technology at the University of Malaya, Malaysia. She received her M.Sc. in Information Systems in 2008 from India and BE (Software Engineering) in 2001 from Iran. She worked in 3MCD and EIC, Iran as a network engineer and participated in several wireless communication projects from 2001 till 2006. She has been working for more than 6 years as a part-time lecturer for the ministry of social affairs, Iran as a technical and vocational trainer. Her main research interests include mobile cloud computing, distributed computing, and ubiquitous computing. She is a member of IEEE society and can be corresponded via zsanaeim@gmail.com or sanaei@ieee.org. For further information, please visit her cyberhome at www.mobilecloudfamily.com
\end{biography}

\begin{biography}{Ejaz Ahmed} was 
born in Gandhian, Mansehra, Pakistan. He did his B.S (Computer Science) from Allama Iqbal Open University, Islamabad, Pakistan. Afterward, he completed his M.S (Computer Science) from Mohammad Ali Jinnah University, Islamabad in 2009. Currently, he is pursuing his PhD Candidature under Bright Spark Program at Faculty of Computer and Information Technology, University Malaya, Kuala Lumpur, Malaysia. He is an active researcher in Mobile Cloud Computing Research Group at Faculty of Computer Science and Information Technology, University Malaya, Kuala Lumpur, Malaysia. His areas of interest include Seamless Application Execution Framework Design for Mobile Cloud Computing, Designing of Channel Assignment and Routing Algorithms for Cognitive Radio Networks.
\end{biography}

\vspace*{-1.5\baselineskip}
\begin{biography}[{\includegraphics[scale=0.46]{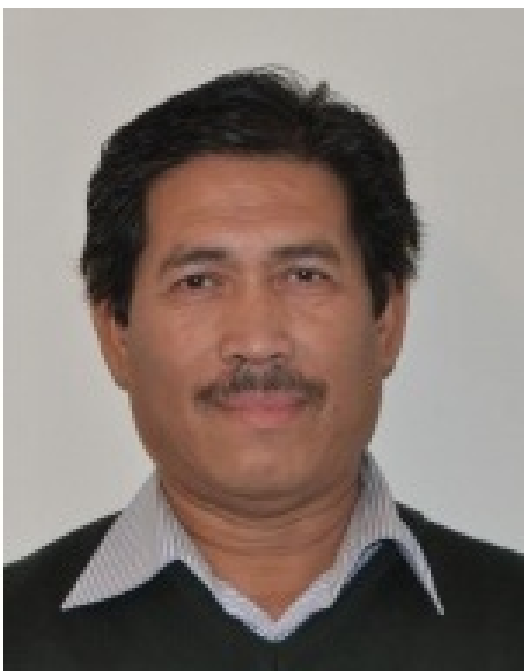}}]{Dr. Abdullah Gani} is Associate Professor of Computer System and Technology at the University of Malaya, Malaysia. His academic qualifications were obtained from UK's universities - bachelor and master degrees from the University of Hull, and Ph.D from the University of Sheffield. He has vast teaching experience due to having worked in various educational institutions locally and abroad - schools, teaching college, ministry of education, and universities. His interest in research started in 1983 when he was chosen to attend Scientific Research course in RECSAM by the Ministry of Education, Malaysia. More than 100 academic papers have been published in conferences and respectable journals. He actively supervises many students at all level of study - Bachelor, Master and PhD. His interest of research includes self-organized systems, reinforcement learning, and wireless-related networks. He is now working on mobile cloud computing with High Impact Research Grant for the period of 2011-2016.
\end{biography}

\vspace*{2\baselineskip}
\begin{biography}[{\includegraphics[scale=1.35]{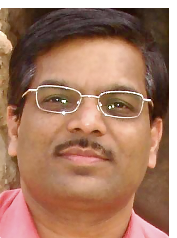}}]{Dr. Rajkumar Buyya} is Professor of Computer Science and Software Engineering, Future Fellow of the Australian Research Council, and Director of the Cloud Computing and Distributed Systems (CLOUDS) Laboratory at the University of Melbourne, Australia. He is also serving as the founding CEO of Manjrasoft, a spin-off company of the University, commercializing its innovations in Cloud Computing. He has authored over 425 publications and four text books including ``Mastering Cloud Computing'' published by McGraw Hill and Elsevier/Morgan Kaufmann, 2013 for Indian and international markets respectively. He also edited several books including ``Cloud Computing: Principles and Paradigms'' (Wiley Press, USA, Feb 2011). He is one of the highly cited authors in computer science and software engineering worldwide (h-index=70, g-index=144, 23000+ citations). Microsoft Academic Search Index ranked Dr. Buyya as the world's top author in distributed and parallel computing between 2007 and 2012. Recently, ISI has identified him as a ``Highly Cited Researcher'' based on citations to his journal papers.

Software technologies for Grid and Cloud computing developed under Dr. Buyya's leadership have gained rapid acceptance and are in use at several
academic institutions and commercial enterprises in 40 countries around the world. Dr.  Buyya has led the establishment and development of key community activities, including serving as foundation Chair of the IEEE Technical Committee on Scalable Computing and five IEEE/ACM conferences.
These contributions and international research leadership of Dr. Buyya are recognized through the award of ``2009 IEEE Medal for Excellence in
Scalable Computing'' from the IEEE Computer Society, USA. Manjrasoft's Aneka Cloud technology developed under his leadership has received ``2010
Asia Pacific Frost \& Sullivan New Product Innovation Award'' and ``2011 Telstra Innovation Challenge, People's Choice Award''. He is currently serving as the foundation Editor-in-Chief (EiC) of IEEE Transactions on Cloud Computing. For further information on Dr. Buyya, please visit his cyberhome: www.buyya.com
\end{biography}

\end{document}